\documentclass[a4paper,11pt]{article}
\pdfoutput=1 

\usepackage{jcappub} 
\usepackage{graphicx}                     

\usepackage[T1]{fontenc} 

\def\be{\begin{equation}}
\def\ee{\end{equation}}
\newcommand{\mst}{{m_{\rm st}}}
\newcommand{\Veff}{{V_{\rm eff}}}

\newcommand{\diag}{\mathrm{diag}}

\newcommand{\Neff}{N_\mathrm{eff}}
\newcommand{\Neffth}{N_\mathrm{eff,th}}

\newcommand{\calN}{\mathcal{N}}

\newcommand{\Hbar}{\widetilde{H}}
\newcommand{\Ebar}{\widetilde{\cal E}}
\newcommand{\tep}{\tilde{\varepsilon}}
\newcommand{\ffd}{f_{\rm FD}}

\newcommand{\fosc}{f_{\rm osc}}

\newcommand{\nns}{n_{\nu_s}}
\newcommand{\dNeff}{\Delta\Neff}
\newcommand{\YP}{Y_{\rm P}}
\newcommand{\YDP}{y_{\rm DP}}
\newcommand{\wb}{\omega_b}
\newcommand{\wc}{\omega_{\rm cdm}}
\newcommand{\sth}{\sigma^{\rm th}}
\newcommand{\dat}{{\rm dat}}
\newcommand{\ssq}{\sin^2\theta_{14}}
\newcommand{\lgGX}{\log_{10}(G_X/G_F)}
\newcommand{\Tnus}{T_{\nu_s}}
\newcommand{\lcdm}{\Lambda {\rm CDM}}
\newcommand{\tre}{\tau_{\rm reio}}
\newcommand{\As}{\ln 10^{10}A_s}
\newcommand{\eV}{~{\rm eV}}
\newcommand{\uH}{{\rm km}{\rm s}^{-1}{\rm Mpc}^{-1}}

\subheader{YITP-SB-18-13}

\title{Cosmological constraints with self-interacting sterile
neutrinos}

\author[a]{Ningqiang Song,}
\author[a,b,c]{M.C. Gonzalez-Garcia,} 
\author[c]{Jordi Salvado}

\affiliation[a]{{\it C.N. Yang Institute for Theoretical Physics,
    State University of New York at Stony Brook, Stony Brook,
    NY 11794-3840, USA}}
\affiliation[b]{\it{Instituci\'o Catalana de Recerca i Estudis
    Avan\c{c}ats (ICREA), Pg. Lluis Companys 23, 08010 Barcelona, Spain}}
\affiliation[c]{\it{Department de Fis\'{\i}ca Qu\`antica i Astrof\'{\i}sica
    and Institut de Ciencies del Cosmos, Universitat de Barcelona,
    Diagonal 647, E-08028 Barcelona, Spain}}

\emailAdd{ningqiang.song@stonybrook.edu}
\emailAdd{maria.gonzalez-garcia@stonybrook.edu}
\emailAdd{jsalvado@icc.ub.edu}
\abstract{In this work we revisit the question of whether
Cosmology can be made compatible with scenarios with light sterile neutrinos,
as invoked to explain the SBL anomalies, in the presence of self-interaction
among sterile neutrinos mediated  by massive gauge bosons. We examine this
proposal by deriving the cosmological predictions of the model in a wide
range of the model parameters including the effective interaction strength
$G_X$, sterile neutrino mass $\mst$ and active-sterile mixings. With those
we perform a statistical  analysis of the cosmological data from BBN, CMB,
and BAO data to infer the posterior probabilities of the sterile
self-interaction model  parameters. BBN mostly provides information
about the effective interaction strength and we find that $\lgGX\geq
{4.0}$ can
describe the primordial abundances at 95\% CL. 
Our analysis of CMB and BAO data show that when allowing a
wide prior for the sterile neutrino mass its posterior is bounded to
$\mst\leq {0.95}\eV$ (95\%~CL) considering CMB data only and
$\mst\leq 0.37\eV$ (95\%~CL) when adding the  BAO information.  So the mass
bounds are slighly relaxed compared with that of a non-interacting sterile
neutrino model but a sterile neutrino mass of 1\,eV is still excluded
at more than $2\sigma$ CL. Conversely if fixing the sterile neutrino mass
and mixing to the values prefered by short baseline data we find that
while that CMB data alone favors the self-interacting scenario,
including the BAO information severly degrades the agreement with
the model. Altogether we conclude then that adding the self-interaction
can alleviate the tension between eV sterile neutrinos and CMB data,
but when including also the BAO results the self-interacting sterile neutrino
model cannot lead to a satisfactory description of the data.}

\begin{document}
\maketitle
\flushbottom
\section{Introduction}
Over the last two decades it has been  solidly established the need to
extend  the leptonic sector of the Standard Model with the addition of 
mass terms for at least two of its three neutrino states, as required
to describe the results of solar, atmospheric, reactor and
long baseline neutrino experiments
(for a review see  ~\cite{GonzalezGarcia:2007ib}).
In this extension, lepton flavors are not symmetries of Nature
~\cite{Pontecorvo:1967fh,Gribov:1968kq} and the minimum joint description of
all these data requires mixing among all the three known neutrinos
($\nu_e$, $\nu_\mu$, $\nu_\tau$), which can be expressed as quantum
superposition of three massive states $\nu_i$ ($i=1,2,3$) with masses
$m_i$ leading to the observed oscillation signals with $\Delta
m^2_{21}=(7.40\pm 0.21) 10^{-4}$ eV$^2$ and
$|\Delta m^2_{3\ell}|=(2.5\pm 0.03)\times 10^{-3}$ eV$^2$ and non-zero
values of the three mixing angles \cite{Esteban:2016qun,nufit-3.0}.

In addition to these well-established results, there remains a set of
anomalies in neutrino data at relatively short-baselines (SBL).  The
first one came from the LSND experiment~\cite{Aguilar:2001ty} which
observed excess events of $\bar{\nu}_\mu\rightarrow \bar{\nu}_e$
oscillations. An excess was also observed in the antineutrino mode of
MiniBooNE~\cite{AguilarArevalo:2010wv} which was consistent with the
results of LSND, while the search for $\nu_\mu\rightarrow\nu_e$
oscillations reported excess only in the low energy
range~\cite{AguilarArevalo:2008rc}. GALLEX~\cite{Anselmann:1994ar,Hampel:1997fc,Kaether:2010ag}
and SAGE~\cite{Abdurashitov:2005tb} detected electron neutrinos
produced by radioactive sources and observed deficit in the $\nu_e$
capture rates. The deficit is known as Gallium anomaly. Furthermore,
some reactor antineutrino measurements also reported deficit in the
$\bar{\nu}_e$ flux~\cite{Mention:2011rk} in the near detectors.  When
interpreted in terms of oscillations, each of these anomalies pointed
out towards a $\Delta m^2\sim {\cal O}({\rm eV}^2)$ and mixing angle
$\theta\sim 0.1$
~\cite{Kopp:2011qd,Giunti:2011gz,Giunti:2011hn,Giunti:2011cp,Donini:2012tt,Conrad:2012qt,Giunti:2013aea,Karagiorgi:2012kw,Kopp:2013vaa,Gariazzo:2013gua,Collin:2016rao}
and consequently could not be explained within the context of the
3$\nu$ mixing described above. They required instead the addition of
one or more neutrino states (what is usually referred to as $3+1$ or $3+N_s$
models)
which must be {\sl sterile}, {\it i.e.} elusive to Standard Model interactions,
to account for the constraint
of the invisible $Z$ width which limits the number of light
weak-interacting neutrinos to be $2.984 \pm
0.008$~\cite{Nakamura:2010zzi}.

Recently, some of these anomalies have been questioned, in particular
with the results of new reactor neutrino experiments from Daya
Bay~\cite{An:2017osx}, NEOS~\cite{Ko:2016owz}, and
DANSS~\cite{Alekseev:2016llm,danss-moriond17,danss-solvay17}, together
with theoretical developments in the calculation of reactor neutrino
fluxes~\cite{Huber:2015ouo,Huber:nuplatformweek}.  But a combined
analysis of Daya Bay data with NEOS and DANSS experiments still allows
for eV sterile neutrinos~\cite{Giunti:2017yid,Dentler:2017tkw} to
explain for disappearance of $\bar\nu_e$.  Mounting tensions arise,
however, to find a consistent description incorporating also the
appearance results from LSND and MiniBoone \cite{Dentler:2018sju}.

Besides the debate on the status of these hints towards light sterile
neutrinos in oscillation experiments, massive sterile neutrino itself
have interesting consequences in Cosmology.
If they have a non-negligible mixing with active neutrinos, light
sterile neutrinos were in thermal equilibrium with the active
neutrinos in the early universe which results in the effective number
of neutrino species $\Neff\simeq 4$ in the 3+1 models. This is in
tension with the precise measurement of primordial abundances produced
in Big Bang Nucleosynthesis (BBN)~\cite{Cyburt:2015mya} and with
Cosmic Microwave Background (CMB) data which generically constraints
$\Neff$ to be close to three, and the total mass of the neutrinos to
be well below eV ~\cite{Ade:2015xua}.

The generic conclusion is that, in order to accommodate the cosmological
observations within the $3+N_s$ scenarios motivated by SBL results, some
new form of physics is required to suppress the contribution of 
the sterile neutrinos to $\Neff$ (see for example \cite{Bergstrom:2014fqa}).
Among others extended scenarios with   
a time varying dark energy component~\cite{Giusarma:2011zq}, entropy
production after neutrino decoupling~\cite{Ho:2012br}, 
very low reheating temperature~\cite{Gelmini:2004ah}, large lepton
asymmetry~\cite{Foot:1995bm,Chu:2006ua,Saviano:2013ktj}, and non-standard
neutrino interactions ~\cite{Bento:2001xi,Dasgupta:2013zpn,Hannestad:2013ana},
have been considered.
All these mechanisms have the effect of diluting the sterile neutrino 
abundance or suppressing its production in the early universe.
In particular the presence of new interactions in the sterile sector have
been proposed to achieve this goal. The interactions via a light pseudoscalar
for the 4th mass state has been studied by Archidiacono {\it et
  al}~\cite{Archidiacono:2014nda,Archidiacono:2016kkh} which lead to a
consistent description of cosmological observations including an upward
revision of
Hubble constant $H_0$. However, in their work the interaction strength
was not mapped into $\Neff$ directly due to the complexity of the
quantum kinetic equations (QKEs). Alternatively in 
Refs.~\cite{Dasgupta:2013zpn,Chu:2015ipa} it was proposed that new
interaction between sterile neutrinos mediated by a new massive gauge
boson $X$ described by a Lagrangian
\be
\mathcal{L}_{\rm int}=
g_X\bar{\nu}_s\gamma_\mu\dfrac{1}{2}(1-\gamma_5)\nu_sX^\mu\,,
\label{eq:Lagrang}
\ee
where $g_X$ is the gauge coupling. When the energy scale is smaller than
the gauge boson mass $M_X$, we can integrate $X$ out to obtain an
effective four sterile neutrino interaction Lagrangian with the
effective coupling $\frac{G_X}{\sqrt{2}}=\frac{g_X^2}{8M_X^2}$. This
new interaction has been studied by Saviano {\it et al} to obtain the
bounds from BBN~\cite{Saviano:2014esa} as well as CMB and baryon
acoustic oscillation (BAO) data~\cite{Forastieri:2017oma}, and most
recently in Ref.~\cite{Jeong:2018yts} from future measurements of the
diffuse supernova background. {In particular in
Refs.~\cite{Saviano:2014esa,Forastieri:2017oma} it is argued that 
this new interaction can reduce $\Neff$ to 2.7 via a late production
of sterile neutrinos. And with a large coupling it can reduce the
free-steaming of neutrinos, which can induce changes in the amplitude
and phase of the acoustic peaks in the CMB
spectrum~\cite{Choi:2018gho}, and avoid the constraint on $\Sigma
m_\nu$ from the measurement of large scale structure. Still, the
study in Ref.~\cite{Forastieri:2017oma} showed that the strong
self-interacting scenario described in
Ref.~\cite{Chu:2015ipa} is excluded by more than 3$\sigma$ (see also
Ref.~\cite{Chu:2018gxk} for a recent discussion)}.

In this work, we have a critical look at the proposal of such
vector-like sterile neutrino self-interactions 
without focusing  on the representative strong-interacting
scenario but rather exploring a large range of effective couplings from
weak to strong. Our aim is to show to what extent this new interaction
can or cannot alleviate the tension between light sterile neutrinos and
Cosmology. Technically our work relies on the formalism developed in
Refs.~\cite{Mirizzi:2012we,Forastieri:2017oma}
to account for the new
sterile neutrino self-interactions in the QKE for the density of the neutrino
ensemble, and in the Boltzmann equations for its perturbations (to which
we introduce some minor improvements) {but
we differ from Refs.~\cite{Mirizzi:2012we,Forastieri:2017oma}
in the assumptions used when obtaining the energy and {{phase space distributions}} of the neutrino ensemble after decoupling (the corresponding value of $\Neff$)  as discussed in Sec.~\ref{sec:forma}.
Furtherore, we extend the study} by  consistently exploring the solutions
obtained as a function of the parameters of the sterile neutrino
self-interacting model. With those at hand we perform a Bayesian
analysis of the cosmological data from BBN, CMB, and BAO data   
to infer the posterior of the sterile self-interaction model  parameters
in Sec.~\ref{sec:results}. We summarize our conclusions in
Sec.~\ref{sec:conclu}. We include an appendix with two short
sections describing the possible dependence of the results on the priors
implied for $\Neff$ and on the range of validity of the study.

\section{Framework}
\label{sec:forma}
Our starting point is a $\Lambda$CDM cosmology extended with one
additional sterile neutrino with mass $\mst$ with non-zero projections
over two of the massive neutrino states (as parametrized by two mixing
angles $\theta_{14}$ and $\theta_{24}$) with self-interactions as in
Eq.~\eqref{eq:Lagrang} with coupling constant $g_X$ mediated by a
massive boson of mass $M_X$ (or equivalently with an effective
self-coupling $G_X$).  For each point in the model parameter space
$\mst$, $\theta_{14}$, $\theta_{24}$, $g_X$, and $G_X$ we first obtain
the effective number of neutrino species at the time of BBN
$\Neff(T_{\rm BBN})$ by solving the QKEs quantifying the neutrino
flavour evolution as described in Secs.~\ref{sec:QKE}
and~\ref{sec:BBNdis}. Subsequently we consistently introduce both the
results of the QKE's relevant for the neutrino background evolution
and the direct effect of the sterile neutrino self-interactions in the
Boltzmann equations for the neutrino perturbations as described in
Sec.~\ref{sec:nuBoltz} and obtain the modified predictions for CMB and BAO
observables.

\subsection{Sterile neutrino production and neutrino flavor evolution}
\label{sec:QKE}

In the early universe the self-interactions in Eq.~\eqref{eq:Lagrang} induce
inelastic collisions among the sterile states with rate 
\begin{equation}
\Gamma_X=\nns\langle \sigma v\rangle \simeq G_X^2T_{\nu_s}^5 \;,
\end{equation}
where $\nns$ and $T_{\nu_s}$ are the number density and temperature of
the sterile neutrinos. Because of the mixing between sterile neutrinos
and active neutrinos these collisions can bring the sterile neutrinos into
thermal equilibrium with the active ones. 

However the self-interactions also induce elastic forward scattering among the
sterile neutrinos which can be parametrized in terms of an MSW-like \cite{Mikheev:1986gs}
effective potential which takes the form ~\cite{Sigl:1992fn}
\be
V_{\rm eff}=-\dfrac{8\sqrt{2}G_X\,p\,\varepsilon_s}{3M_X^2}\,,
\ee
where $p$ is the momentum of the sterile neutrino and
$\varepsilon_s$ is its energy density. As discussed in
Refs.~\cite{Dasgupta:2013zpn,Chu:2015ipa}  by introducing this effective
potential the in-medium mixing angle between active and sterile neutrinos
deviates from its vacuum value by
\begin{equation}
\sin^22\theta_m\,=\,\dfrac{\sin^22\theta_0}{\left(\cos 2\theta_0\,+
  \,2E/\mst^2\,\Veff\right)^2\,+\,\sin^22\theta_0}\,,
\label{eq:sinthetam}
\end{equation}
So if $\Veff\gg \fosc= \mst^2/(2E)$ before neutrinos decoupling from
the primordial plasma, the in-medium mixing angle is highly suppressed
and sterile neutrino production is deferred until after decoupling.
By comparing the interaction rate with the expansion rate, and the
self-interaction potential with the oscillation frequency, one finds that this
mechanism can work for strong enough interactions.

However it was {argued in Ref.~\cite{Mirizzi:2014ama}  that
entropy should be conserved after decouling so the
entropy possessed by the three active neutrino species had to be shared
with the sterile neutrinos. This assumption leads to a reduction in the total
neutrino energy density which implies that $\Neff$ could be as low as
2.7 instead of 4.  A value which would be  too low when confronted
with data ~\cite{Forastieri:2017oma} unless some additional mechanism besides
mixing was invoked
to produce the sterile neutrinos \cite{Chu:2015ipa}. As we discuss below
we depart from the entropy conservation assumption employed in
Ref.~\cite{Mirizzi:2014ama} when evaluating $\Neff$ {{which}} may affect the
conclusions about this scenario in the strong coupling regime as
we describe next.}


The quantitative determination of $\Neff$ in
scenarios with light sterile neutrinos which are brought in to
equilibrium by their mixing with the standard three active neutrinos
requires to evaluate the time evolution of their energy density. To do
so we can use the quantum kinetic equations (QKEs) of the 3+1 neutrino
ensemble described in Ref.~\cite{Mirizzi:2012we}. For sake of
completeness we briefly summarize them here {with the inclusion of 
two minor improvements: considering a separate treatment of
neutrino and photon temperature, and an improved determination of
the electron energy density.}

The flavour evolution of the 3+1 neutrino ensemble is described in terms
of a 4x4 density matrix $\varrho(p,t)$  whose  evolution is governed by the QKEs 
\begin{equation}
{\rm i}\,\frac{d\varrho}{d t} =[\Omega,\varrho]+ C[\varrho]\, ,
\label{eq:drhodt}
\end{equation}
where $C[\varrho]$ represents the collision terms while the
oscillation and in-medium potential terms corresponding to charged
current ($CC$) interactions with the background electrons and the
neutral current ($NC$) interactions with the background neutrinos are
\begin{equation}
\Omega=\dfrac{1}{2p}U^\dagger
M^2U+\sqrt{2}G_F
\left[-\dfrac{8p}{3}\left(\dfrac{{\cal E}_l}  {M_W^2}
  +\dfrac{{\cal E}_\nu}{M_Z^2}\right)\right]
+\sqrt{2}G_X\left[-\dfrac{8p}{3}\dfrac{{\cal E}_s}{M_X^2}\right]\,.
\label{eq:matterpotential}
\end{equation}
$M^2=\diag (0,\Delta m_{21}^2,\Delta m_{31}^2, m_{\rm st}^2)$
(we assume the sterile neutrino mass is much larger than the active
ones), and $M_W$ and $M_Z$ are the masses of $W$ and $Z$ bosons
respectively. $U=R_{34}R_{24}R_{23}R_{14}R_{13}R_{12}$
($R_{ij}$ representing a rotation of angle $\theta_{ij}$ in the
${ij}$ plane) is the neutrino mixing matrix.

In what follows we fix the oscillation parameters for the three
active neutrinos $\Delta m^2_{21}$, $\Delta m^2_{31}$, $\theta_{12}$ ,
$\theta_{13}$, and $\theta_{23}$ to the best fit for normal ordering
from the global oscillation analysis in NuFIT 3.0
~\cite{Esteban:2016qun,nufit-3.0}.  In Eq.~\eqref{eq:matterpotential}
${\cal E}_l=\diag(\varepsilon_e,0,0,0)$, ${\cal E}_\nu$, and ${\cal
  E}_s=\diag(0,0,0,\varepsilon_s)$ are $4\times 4$ matrices containing
the energy density of the electrons, active neutrinos (which is in
general non-diagonal with non-zero entries in the upper $3\times 3$
sector), and sterile neutrinos respectively.

Equation ~\eqref{eq:drhodt}, though well-defined, is extremely
computationally demanding due to the momentum dependence of the
density matrix, especially in the case of three active plus one
sterile neutrino species. To retain the main features of the flavor
evolution within a reasonable amount of computing time, we still
resort to the average momentum approximation as described
in~\cite{Mirizzi:2012we}. In this approximation, ones remove the
momentum dependence in the equations by assuming
\begin{equation}
  \varrho(x,y)\longrightarrow \ffd(y)\rho(x) \;,
\label{eq:rhoav}  
\end{equation}
where $\ffd$ Fermi-Dirac distribution function for the neutrinos
and for convenience
we have introduced the dimensionless variables 
\begin{equation}
x\equiv ma,\quad y\equiv pa,\quad z_\gamma\equiv T_\gamma a,\quad
z_\nu\equiv T_\nu a,
\end{equation}
where we take the arbitrary mass scale $m$ to be 1~MeV. {This assumes neutrinos are featured by grey body Fermi-Dirac distributions. We shall use Eq.~\eqref{eq:rhoav} later to restore the momentum dependent phase space distribution of neutrinos.} We stress that
we have added $z_\nu$ to trace the difference between neutrino and
photon temperatures at or after the time of $e^+e^-$ annihilation.
Since we are solving the equations below $\mu^+\mu^-$ annihilation,
always $T_\nu\propto 1/a $ so $z_\nu$ is a constant (hence the $\ffd$ in
Eq.~\eqref{eq:rhoav} is only a function of $y$). In what follows we normalize
the scale factor to  $a(t)=1/T_\nu$ so that $z_\nu(x)$ is always 1.
However $z_\gamma$ will evolve.  Indeed we can solve
for $z_\gamma(x)$ from the conservation of stress-energy tensor
(see Eq.~(15) in~\cite{Esposito:2000hi} for
a detailed treatment).
The solution does not depend on the details of the neutrino flavor evolution,
so this $z_\gamma(x)$ is precomputed as a known function before solving the QKEs.

Altogether one finds
\begin{eqnarray}
i\dfrac{d\rho}{d x}=&+&\dfrac{x^2}{2m^2\Hbar}\langle
\dfrac{1}{y}\rangle\left[U^\dagger M^2
  U,\rho\right]+\dfrac{\sqrt{2}G_F m^2}{x^2
  \Hbar}\left[\left(-\dfrac{8 \langle y\rangle
    m^2}{3x^2M_W^2}\Ebar_l-\dfrac{8 \langle y\rangle
    m^2}{3x^2M_Z^2}\Ebar_\nu\right),\rho\right]\nonumber\\ &+&\dfrac{\sqrt{2}G_X
  m^2}{x^2 \Hbar}\left[-\dfrac{8 \langle y\rangle
    m^2}{3x^2M_X^2}\Ebar_s,\rho\right]+\dfrac{x\langle
  C[\rho]\rangle}{m\Hbar}\,,
\label{eq:avedensitymatrix}
\end{eqnarray}
where in the potentials
\begin{eqnarray}
  \Ebar_l&=&\diag  (\tep_e,0,0,0)\,,\\
  \Ebar_\nu&=&\dfrac{2}{2\pi^2}\int_0^\infty
d y y^3
G_s\varrho(x,y)G_s=\dfrac{7}{8}\dfrac{\pi^2}{15}G_s\rho(x)G_s\,,\\
\Ebar_s&=&\dfrac{2}{2\pi^2}\int_0^\infty
d y y^3
G_{sX}\varrho(x,y)G_{sX}=\dfrac{7}{8}\dfrac{\pi^2}{15}G_{sX}\rho(x)G_{sX}\,.
\end{eqnarray}
In the above equations $G_s=\diag(1,1,1,0)$ and $G_{sX}=\diag(0,0,0,1)$
contain the dimensionless coupling constants for active and sterile neutrinos,
respectively. We have introduced  the normalized Hubble parameter as
\begin{equation}
\Hbar\equiv \dfrac{x^2}{m}H=\dfrac{m}{M_{\rm
    Pl}}\sqrt{\dfrac{8\pi\tep(x,z_\gamma(x),z_\nu(x))}{3}}\,,
\end{equation}
where $M_{\rm Pl}$ is the Planck mass and the comoving total energy
density is defined as $\tep\equiv \varepsilon(x/m)^4$ with
$\varepsilon=\tep_e+{\rm Tr}(\Ebar_\nu+\Ebar_s)$ is  the total energy
density. If we define the ``reference'' cosmological model as the $\lcdm$ model
plus three active neutrinos, in this model neutrinos keep in thermal
equilibrium with the plasma until decoupling. After $e^+e^-$
annihilation, $T_\nu=(4/11)^{1/3}T_\gamma$. The number density of a
neutrino species in the reference model is denoted by $n_\nu^*$. In
terms of the average momentum approximation, the diagonal entries in
the density matrix denote the number density of active or sterile
neutrinos normalized to $n_\nu^*$.

Also when computing $\tep_e$ we use
\begin{equation}
  \tep_e=\dfrac{2}{\pi^2}\int_0^\infty d y y^3  {\ffd}_e(y,z_\gamma(x)),
  \label{eq:epsie}
\end{equation}
with ${\ffd}_e=g_e/\exp(\sqrt{y^2+x^2m_e^2/m^2}/z_\gamma+1)$ which is
slightly different from  from Eq.~(18) in Ref.~\cite{Mirizzi:2012we} in that we keep
the electron mass so the energy density of electrons quickly approaches
0 during annihilation.

In Eq.~\eqref{eq:avedensitymatrix}, $\langle C[\varrho]\rangle$ is
the momentum average of active and sterile collision terms expressed
as~\cite{Chu:2006ua,Mirizzi:2012we}
\begin{eqnarray}
\langle C_\nu[\varrho]\rangle &=&
-\dfrac{i}{2}G_F^2\dfrac{m^5}{x^5}(\{S^2,\rho-I\}-2S(\rho-I)S+\{A^2,\rho-I\}+2A(\rho-I)A\,,\nonumber\\ \langle
C_s[\varrho]\rangle &=&
-\dfrac{i}{2}G_X^2\dfrac{m^5}{x^5}(\{S_X^2,\rho-I\}-2S_X(\rho-I)S_X)\,,
\end{eqnarray} 
where the active neutrino scattering and annihilation matrix
$S=\diag(g_s^e,g_s^\mu,g_s^\tau,0)$ and
$A=\diag(g_a^e,g_a^\mu,g_a^\tau,0)$ with
$(g_s^e)^2=3.06$, $(g_a^e)^2=0.50$, $(g_s^{\mu,\tau})^2=2.22$,
and $(g_a^{\mu,\tau})^2=0.28$ ~\cite{Chu:2006ua} while 
for sterile neutrinos $S_X=\diag(0,0,0,1)$. We always work in the
approximation $T_\nu<M_X$ so  annihilation terms are neglected
we lose the details of the phase-space distribution, so we always
assume all neutrino species share the same temperature when solving
the QKEs. This assumption is robust since even if there are small
differences between the temperature of the different neutrino species, the
in-medium potentials 
are only corrected by factors of ${\cal O}(1)$ and the solution of the
QKEs would be barely changed.

\begin{figure}[!ht]
\centering \includegraphics[width=0.9\textwidth]{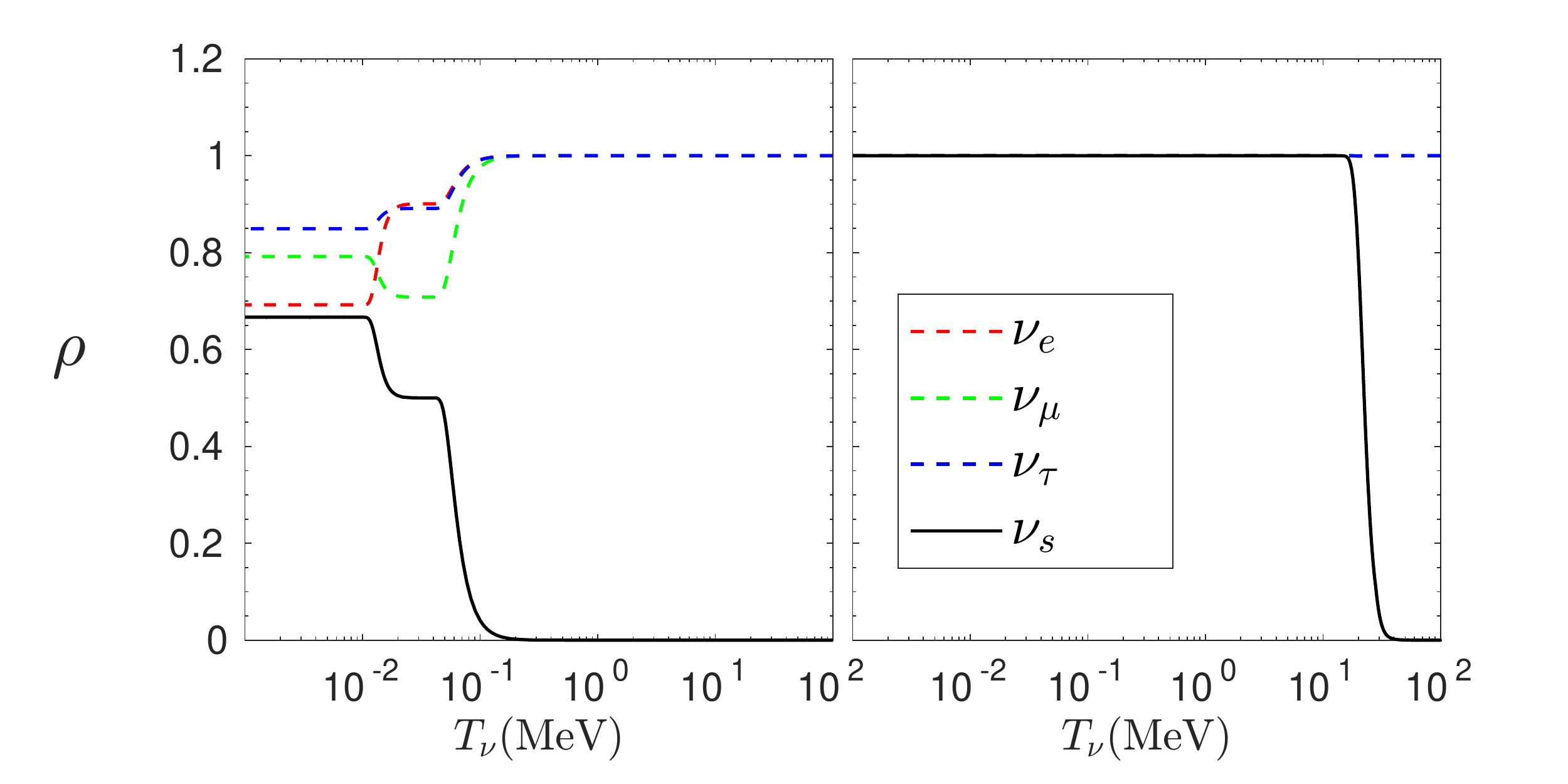}
\caption{The number density $\rho_{\alpha\alpha}$ of active and sterile
  neutrino species as
  a function of the neutrino temperature. The left panel shows the flavor
  evolution for $G_X=10^{10}G_F$, $g_X=0.1$, $\mst=1~{\rm eV}$ and
  $\sin^2\theta_{14}=0.1$, $\sin^2\theta_{24}=0.003$.
  The right panel corresponds to standard
  4$\nu$ evolution without any new interactions.}
\label{fig:flavorevo}
\end{figure}

We solve these equations for an array of values of the model parameters
$\mst$, $\theta_{14}$, and $\theta_{24}$ (for simplicity we neglect mixing
$\theta_{34}$), $g_X$ and $G_X$ (or equivalently $M_X$) and obtain
as a solution $\rho(T_\nu)$.  As illustration of the output of the QKEs
we show in
Fig.~\ref{fig:flavorevo} an example of the neutrino flavor evolution with
parameters such that the new interaction is much
stronger than electroweak interactions. As one would expect, sterile
neutrinos are produced only when the temperature is well below 1~MeV and
all four neutrino species get to similar number densities after
thermalization. For comparison we show in
the right panel the neutrino flavor evolution in the absence
of new interactions. In this case sterile neutrinos get immediately
thermalized with the active neutrinos.
It clearly shows that the self-interaction among sterile neutrinos, when strong
enough, is capable of postponing their production 
until after active neutrino decoupling.

\begin{figure}[!ht]
\centering \includegraphics[width=0.6\textwidth]{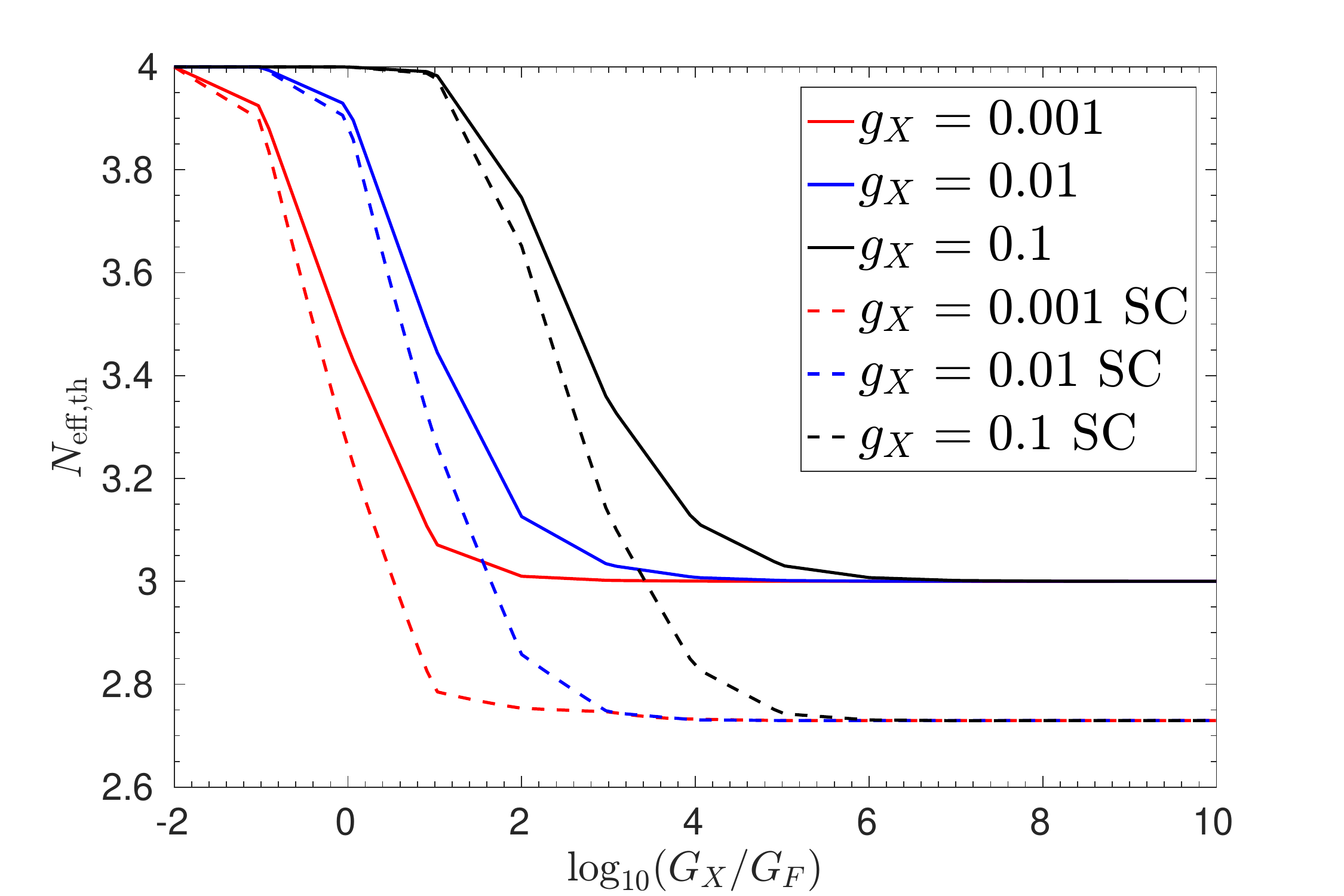}
\caption{$\Neff$ as a function of effective coupling $G_X$. The red,
  blue and black curves correspond to $g_X=$0.001, 0.01, 0.1,
  respectively. Other parameters are fixed to $\mst=1~{\rm eV}$ and
  $\sin^2\theta_{14}=0.01$, and $\sin^2\theta_{24}=0.003$.
 { Full lines correspond to the values obtained with Eq.~\eqref{eq:Neffgood}
while dashed lines are obtaied under the entropy-conservation-after-decoupling
  assumption. Eq.~\eqref{eq:Neffsol}.}}
\label{fig:NeffofGX}
\end{figure}

{Next we need to obtain the value of $\Neff$ corresponding to the
output neutrino matrix density. To this point it is important to notice
that technically, under the average momentum approximation
\begin{equation}
\Neff={\rm Tr}\rho(T)=\sum_{\alpha=1}^{4}\rho_{\alpha\alpha}. 
\label{eq:Neffgood}
\end{equation}   
However, as mentioned above, in Refs.~\cite{Mirizzi:2014ama, Forastieri:2017oma}
$\Neff$ was obtained under the assumption of  total entropy and
number conservation in the neutrino sector after decoupling.
This assumption implies that the entropy initially shared by the three active
neutrino species  would be later shared by the sterile neutrinos as well
through thermalization, which would lead to a reduction in the total  neutrino
energy density, ie in $\Neff$. For instantaneous
decoupling at $T=1$ MeV and Fermi-Dirac distribution with zero chemical
potential for the neutrinos during their flavor evolution this assumption
leads to 
\begin{equation}
 \Neff(T<1\;\, {\rm MeV})=\left(
 \frac{\displaystyle {\rm Tr}\rho({\rm 1\,MeV})}
    {\displaystyle {\rm Tr}\rho(T)} \right)^{\frac{4}{3}}
{\displaystyle \sum_{\alpha=1}^4 (\rho_{\alpha\alpha}(T))^{\frac{4}{3}}}\,.
 \label{eq:Neffsol}
\end{equation}


We show in Fig.~\ref{fig:NeffofGX} $\Neff$ obtained using
Eq.~\eqref{eq:Neffgood} or Eq.~\eqref{eq:Neffsol}  
as a function of $G_X$ for a temperature well below thermalization,
$\Neffth$. As expected for very small $G_X$ early thermalization leads
to $\Neff\simeq 4$ in either case  but  for very large $G_X$, late
thermalization leads to $\Neff\simeq 2.7$
when using  Eq.~\eqref{eq:Neffsol} while $\Neff\simeq 3$ for
Eq.~\eqref{eq:Neffgood}. This last result is physically sensible because
an isolated isotropic gas of effectively massless particles has
equation of state $\omega=1/3$ irrespective of its internal dynamics
{ if the interaction range is smaller that the typical
  distance between the particles. This is a good approximation}
so $\Neff$ is always determined by the total energy density {
  of} the system at the time of decoupling and it is therefore
$\gtrsim 3$ 
\footnote{We thank the anonymous referee of the first version of this
article for pointing out to us this missunderstanding in the literature.}.
It is also the case that, even without impossing entropy conservation
after decoupling, when
working under the average momentum approximation one can find unphysical
values of $\Neff$ if one assumes instantaneous decoupling and that the neutrinos
follow Fermi-Dirac distribution {{with}} zero-chemical potential when relating their number and energy densities, in particular when sterile
neutrino production begins right around decoupling.  Altogether we
find that under the average momentum approximation using
Eq.~\eqref{eq:Neffgood} represents our best estimate of a physically
meaningful $\Neff$ for the neutrino ensemble.  In the following
sections, and when relevant, we will discuss how the results of the
analysis vary when using the assumptions behind {{Eq.~\eqref{eq:Neffsol}}}
(which we will label as SC for ``entropy conservation'') vs
{{Eq.~\eqref{eq:Neffgood}}} for which  no label will be included.
}


\subsection{Effects in  Big Bang Nucleosynthesis}
\label{sec:BBNdis}
The $\Neff$ and neutrino number density we have obtained in the previous
section have direct effects on the primordial abundances. In the early
universe when temperature is much higher than 1~MeV, the primordial
plasma consisting of photons, neutrinos, electrons and baryons is in
thermal equilibrium with the ratio of the number density of neutrons
and protons given by
\be
\left(\dfrac{n_n}{n_p}\right)_{\rm
  eq}=\left(\dfrac{m_n}{m_p}\right)^{3/2}e^{-Q_{np}/T}\,,
\ee where
$Q_{np}=m_n-m_p=1.3~{\rm MeV}$.
Because of the small mass difference the neutron-to-proton ratio decreases
drastically when the temperature drops below 1~MeV. Neutrons and protons are balanced mainly though beta
decay and inverse beta decay, i.e.
\begin{eqnarray}
n+\nu_e&\leftrightarrow &p+e^-\,\nonumber\\ n+e^+&\leftrightarrow &
p+\bar{\nu}_e\,.
\end{eqnarray}
The two interaction rates are equal and can be approximated by
$\Gamma_{np}\propto G_F^2T^5$ provided that $n_e=2n_{\nu_e}$ since
they share the same matrix element. In case neutrino number density
deviates from this relation, the neutron-proton conversion rate
\be
\Gamma_{np}\propto (1+\rho_{ee})T^5\,, 
\ee
At about 1~MeV the
relativistic degree of freedom $g_*=10.75+\dNeff\,$ where
$\dNeff=\Neff-3.046$. In the radiation dominated era, the total energy density $\rho_{\rm tot}\propto g_*T^4$, and the Hubble rate $H\propto
g_*^{1/2}T^2$. Neutrons freeze out from the plasma when the
neutron-proton conversion rate becomes comparable to the the Hubble
rate, and we can find the freezing-out temperature 
\be
T_f\propto \left(\dfrac{g_*^{1/2}}{1+\rho_{ee}}\right)^{1/3}\,.
\label{eq:neutrondec} 
\ee
It shows that both $\Neff$ and electron neutrino number density can
affect the time of neutron decoupling. A larger $\Neff$ can increase
the expansion rate of the universe, which increases the freezing-out
temperature; on the other hand, larger number density of electron
neutrinos tend to increase the neutron-proton conversion rate,
postponing the time of decoupling. The neutron-to-proton ratio
$n_n/n_p$ is roughly frozen after neutron decoupling until the
temperature drops below 0.1~MeV, when the synthesis of nuclei
begin. Almost all the remaining neutrons are bounded in the nuclei,
thus the neutron-to-proton ratio at decoupling is of crucial
importance in the determination of the abundances of primordial
elements. We can incorporate the effect of $\nu_e$ number density by
assuming $\rho_{ee}=1$ but modifying $g_*$ to keep $T_f$ unchanged. It
is straightforward to find the modified $\dNeff$ to be
\be
\dNeff'=\dfrac{4}{7}\left[\dfrac{43+7\dNeff}{(1+\rho_{ee})^2}-10.75\right]\,.
\ee
This treatment is similar to
that of Dolgov {\it et al.}~\cite{Dolgov:2003sg}. Since $\Neff$ and
$\rho_{ee}$ are rather insensitive to the small change in $T_f$, while the
neutron-to-proton depends on $T_f$ exponentially, we use the $\Neff$
and $\rho_{ee}$ at 1~MeV to determine the ratio $n_n/n_p$.

\subsection{$\nu_s$ self-interaction in neutrino perturbations}
\label{sec:nuBoltz}
Next we would like to see quantitatively how the new interactions can
affect the predictions of CMB and large scale structure (LSS)
data. New interactions add collision terms to the Boltzmann equation
and make the solution quite complicated. However, it has been shown in
Ref.~\cite{Oldengott:2017fhy} that an exact description of neutrino
interactions is quantitatively equivalent to the relaxation time
approximation where the collision term can be approximated
by~\cite{Hannestad:2000gt,Forastieri:2017oma}
\be
\dfrac{1}{f_0}\dfrac{\partial f}{\partial
  \tau}=-\dfrac{\calN}{\tau_\nu}\,, \ee where $\tau_\nu=({a}\nns\langle
\sigma v\rangle)^{-1}$ is the mean conformal time between
collisions and $\calN$ is the phase space perturbation of neutrinos. Since $\langle \sigma v\rangle\simeq G_X^2\Tnus^2$, we
have
{
\be
\tau_\nu^{-1}=\dfrac{3}{2}\dfrac{\zeta(3)}{\pi^2}{a}G_X^2 {T_\nu^5 \rho_{ss}}\,,
\ee
where $\rho_{ss}$ is obtained from the solution of QKEs and
$T_\nu=\left(\frac{4}{11}\right)^{1/3} T_\gamma$
is the neutrino temperature
in the standard cosmology} \footnote{{This is also at difference with the
expression employed under the  entropy-conservation-after-decoupling assumption
in Ref.~\cite{Forastieri:2017oma} which translates into the same Fermi-Dirac
distribution for all neutrino especies with a reduced termperature
$T_\nu=\left(\frac{3}{11}\right)^{1/3} T_\gamma$.}}.

In the Synchronous gauge, the neutrino Boltzmann equation can be written as
\be 
\dfrac{\partial
  \calN_i}{\partial \tau}+i\dfrac{q}{\epsilon}(\vec{k}\cdot
\hat{n})\calN_i+\dfrac{d\ln f_0}{d\ln
  q}\left[\dot{\eta}-\dfrac{\dot{h}+6\dot{\eta}}{2}(\vec{k}\cdot
  \hat{n})^2\right]=-\Gamma_{ij}\calN_j,.  
  \ee 
  where $i=1,4$
represent the mass eigenstates, $q=ap$ and $\eta$ and $h$ are
gravitational potentials (not to be confused with Hubble
parameter). $\Gamma_{ij}$ is defined in mass basis as
$\Gamma_{ij}=U\diag(0,0,0,1)U^\dagger\cdot \tau_\nu^{-1}$

We can expand the perturbation $\calN$ as in Legendre
series~\cite{Ma:1995ey} and rewrite the Boltzmann equation as
(below we denote by ``dot'' the derivative with respect to proper time)  
\begin{eqnarray}
  \dot{\calN_{i,0}}&=&-\dfrac{qk}{\epsilon}\calN_{i,1}+\dfrac{1}{6}
  \dot{h}\dfrac{d\ln
  f_{i,0}}{d\ln
    q}\,,\\ \dot{\calN_{i,1}}&=&\dfrac{qk}{3\epsilon}(\calN_{i,0}-2\calN_{i,2})\,,\\
  \dot{\calN_{i,2}}&=&\dfrac{qk}{5\epsilon}(2\calN_{i,1}-3\calN_{i,3})
  -(\dfrac{1}{15}\dot{h}+\dfrac{2}{5}\dot{\eta})\dfrac{d\ln
  f_{i,0}}{d\ln
    q}-\Gamma_{ij}\calN_{j,2}\,,\\
  \dot{\calN_{i,l}}&=&\dfrac{k}{(2l+1)}\dfrac{q}{\epsilon}
  \left[l\calN_{i,(l-1)}-(l+1)\calN_{i,(l+1)}\right]-\Gamma_{ij}\calN_{j,l}\,,\quad
l\geq 3\,,
\end{eqnarray}
{where $f_{i,0}=\ffd  \sum_{\alpha} |U_{\alpha i}|^2\rho_{\alpha\alpha}$,  
with $\ffd$ a Fermi-Dirac distribution of temperature
$T_\nu=\left(\frac{4}{11}\right)^{1/3} T_\gamma$  for the four neutrinos
and again differs from the $f_0$ in Ref.~\cite{Forastieri:2017oma}.} {As we mentioned before, this grey body Fermi-Dirac distribution is the assumption of average momentum approximation which yields the most physical $\Neff$ and we restore it here using Eq.~\eqref{eq:rhoav}.}
In order to account for these effects in the analysis of CMB and BAO
data we have solved the new collisional Boltzmann equations in a modified
version of the Boltzmann code CLASS~\cite{Lesgourgues:2011re} for an array
of values of the model parameters.
Furthermore to account for the effects on the background equations we
have to include also the modified energy density and pressure of the
neutrinos at the starting of the background evolution in CLASS. {These are obtained from the solutions of the QKE's described in the previous section with the average momentum approximation. Under this approximation neutrinos are featured by grey body Fermi-Dirac distributions with normalization factors $\sum_\alpha|U_{\alpha i}|^2 \rho_{\alpha\alpha}$. These distributions are introduced in the CLASS code to infer the background quantities including energy density and pressure}
\footnote{Clearly this assumes that neutrino thermalization, ie the
  temperature for which the QKE's solutions become constant, occurs
  well before the relevant times for the perturbation evolution
  equations. { This  is a very good approximation in all the
    parameter space.}}  

Qualitatively this background effects can be understood in terms of a
modification of ${\Neff}$   at CMB times though technically we do not use the
variable $\Neff$ when solving the CLASS equations. In any case to help
the discussion in what follows we mention an effective $\Neff$ at
CMB times, $\Neffth$,  obtained from Eq.~\eqref{eq:Neffgood} {(or
Eq.~\eqref{eq:Neffsol} when relevant)}
for temperatures well below thermalization and well above the non-relativistic
transition for the sterile neutrino.

\begin{figure}[!htb]
  \centering
\includegraphics[width=0.45\textwidth]{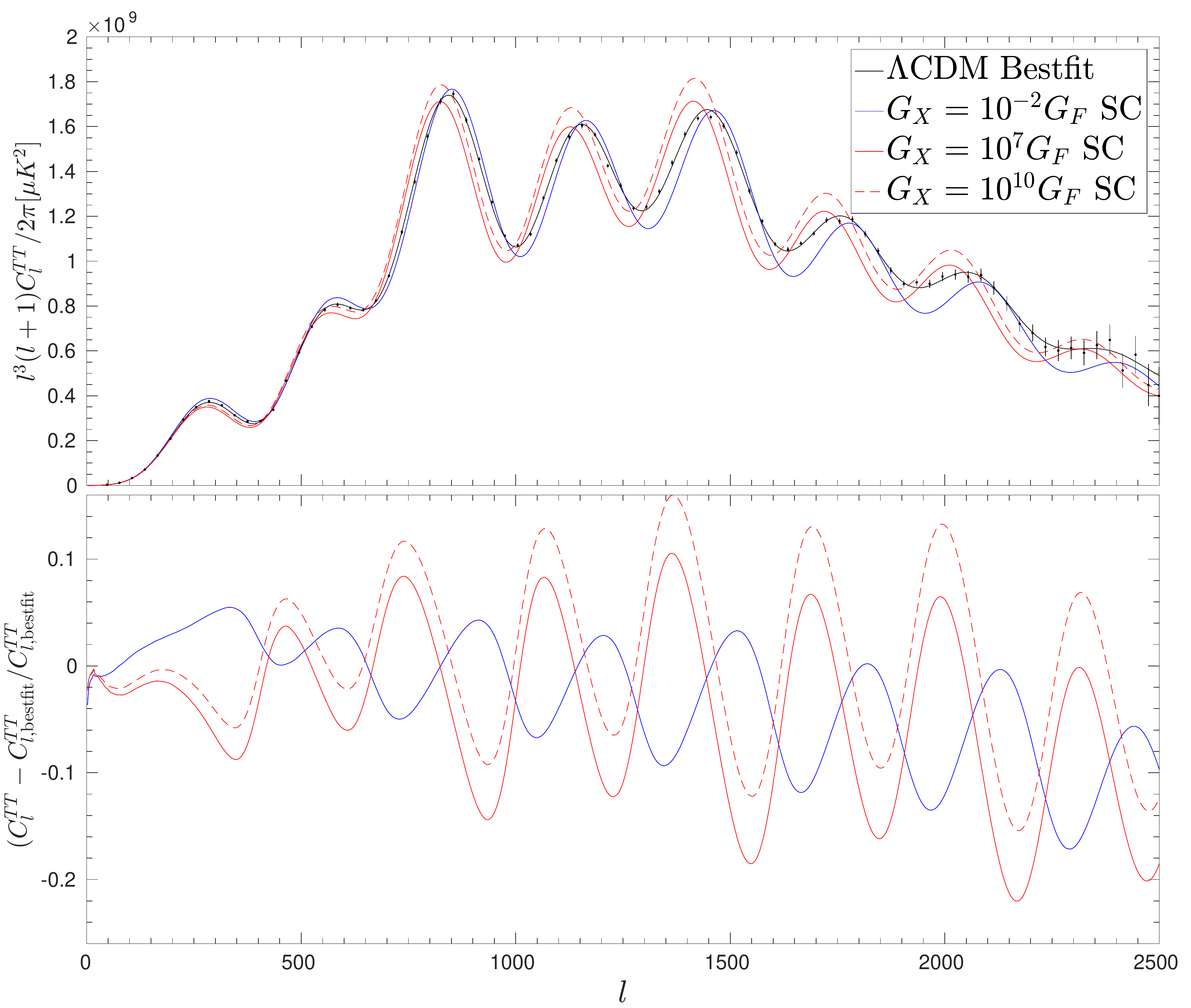}
\includegraphics[width=0.45\textwidth]{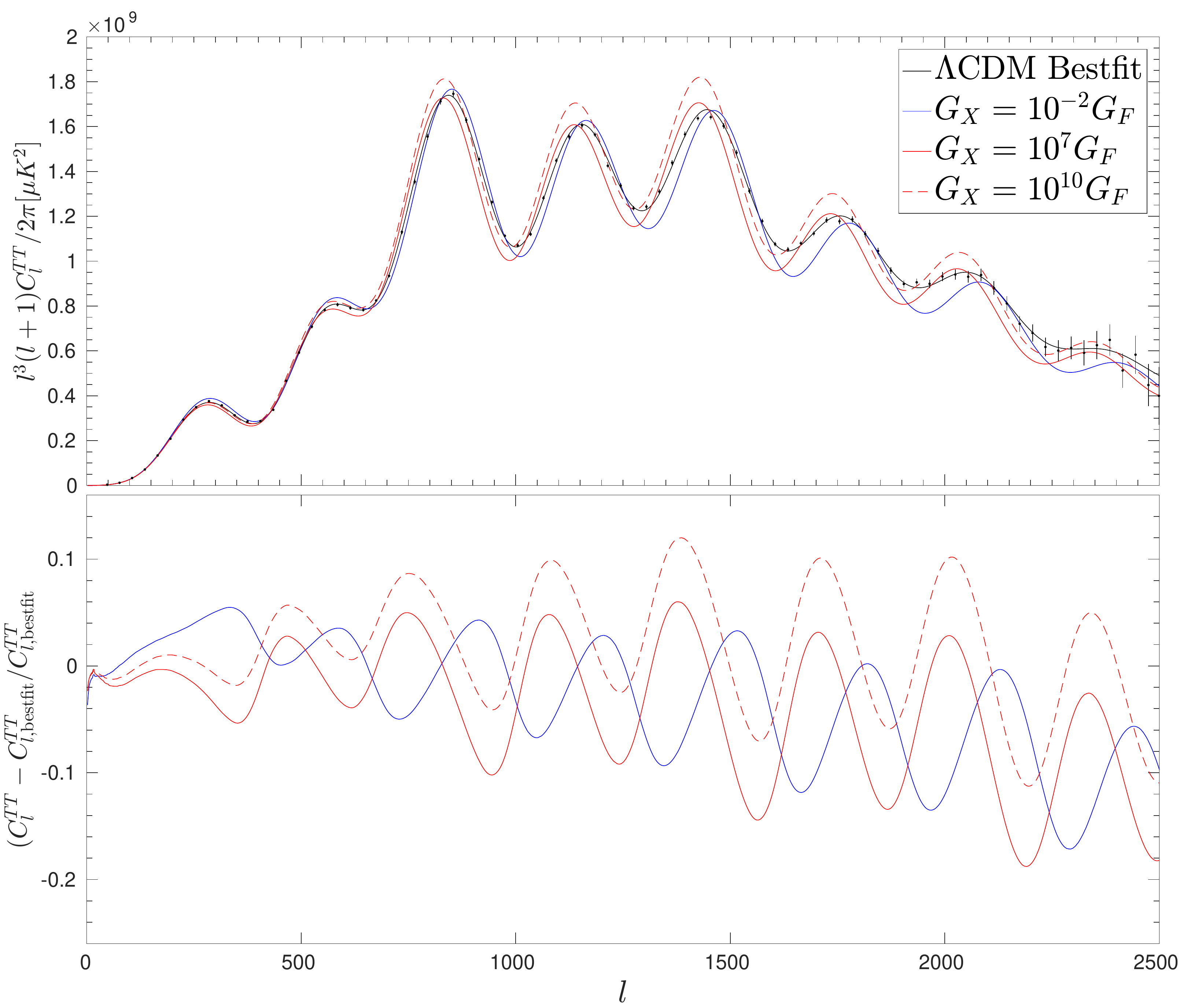} 
\caption{CMB temperature power spectrum. The predicted spectrum in $\lcdm$
for the best fit of Planck TT+lowP~\cite{Ade:2015xua}
($\wb=0.02222$, $\wc=0.1197$,
$H_0=67.31\uH$, $\tre=0.078$, $\As=3.089$, and $n_s=0.9655$)
(i.e. the size of the sound horizon at recombination
$100\theta_{\rm MC}$ has been adjusted to fix $H_0=67.31\uH$ in these curves,
and this is so to illustrate the effect of new interaction on the size
of the sound horizon)
is shown in black along with error bars
from Planck 2015 data. Colored lines correspond to $\lcdm$ models
with one sterile neutrino species and new interactions. Cosmological
parameters are the same as the best fit of Planck TT+lowP and the
implicit parameters for new interactions are the same as
Fig.~\ref{fig:flavorevo} ($g_X=0.1$, $\mst=1~{\rm eV}$, and
$\sin\theta_{14}=0.1$). {
In the left panels the results are obtained under the assumption of entropy
conservation after decoupling employed
Ref.~\cite{Forastieri:2017oma} while in the right we show our results.}
The solid blue, solid red and dashed red lines show different
interaction strength corresponding to {$G_X=10^{-2}G_F\,
(\Neffth=4$ in both left and right panels), $G_X=10^{7}G_F\,$
(i.e. $\Neffth=2.7$ in the left panel while
$\Neffth=3$ in the right),
$G_X=10^{10}G_F$ (i.e. $\Neffth=2.7$ in the left panel while
$\Neffth=3$ in the right),
  respectively}. The lower panel displays the relative difference
  between models with different interaction strength and the best-fit
  $\lcdm$ model.}
\label{fig:Cl}
\end{figure}

For illustration we plot in Fig.~\ref{fig:Cl} the predicted CMB power
spectrum for set of model parameters.
{For comparison we show in the left panels the
  results obtained under the assumption of entropy conservation after
  decoupling employed
Ref.~\cite{Forastieri:2017oma} while on the right we show our results.
In either case,} we have set the collision terms to zero
in the equations of monopole and dipole to ensure particle number and
momentum conservation. In case $\Gamma_{ij}>\cal{H}$, the quadrupole
(related to the anisotropic stress of neutrinos) and higher multipoles
are suppressed. So in the presence of the interactions the power in
higher multipoles are now transferred to the density and velocity
fluctuations which in turn contribute to the total gravitational
source and enhance the amplitude of the CMB fluctuations that entered
the horizon before recombination. This enhancement is clearly seen in
the figures when we compare the red line ($G_X=10^{7}G_F$) with the
dashed red line ($G_X=10^{10}G_F$).

We also notice that the location of the peaks is shifted in the new
interaction models compared with the $\lcdm$ best fit. This is
the effect associated with the modification of the neutrino
background phase space distributions and can be understood qualitatively
as the modification of $\Neff$ and it is therefore most different
between the right and left panels. For models
with strong self-interactions for which at the CMB times
$\Neff<\Neffth \leq 3 \; (2.7)$  in the right (left) panels,
and the peaks move to the left as expected from
the earlier time of mater-radiation equality ($\Neff<\Neffth$ 
because for CMB observables the contribution of the sterile neutrino
to the radiation energy density is further reduced
because they become partially non-relativistic during recombination).
The shift is consequently more pronounced for the panels on the left. 
The solid blue line
($G_X=10^{-2}G_F$), on the other hand, shows the opposite behaviour
since for these weak interactions the resulting $\Neff$ is always larger than
3. Indeed, as a cross check, we have verified explicitly that the blue
line can exactly mimic the behavior of a sterile species without
interactions. The new interaction in this case is too weak to be
significant in the thermalization of sterile neutrinos.

\section{Data analysis: results}
\label{sec:results}

In this section we show the results of confronting the 
sterile self-interaction model with BBN, CMB and BAO data.
Technically to obtain the predictions for these observables as a function of
the model parameters we interface the solutions
of QKEs  with MultiNest \cite{Feroz:2007kg,Feroz:2008xx,Feroz:2013hea} 
or with the modified CLASS by interpolating the
tabulated solutions in the model parameter space and feed them to the codes.

Our aim is to explore as large model parameter space as possible but solving
the QKEs and running CLASS is time demanding. So as a compromise, we allow the
parameters to vary within the range $G_X=10^{-2}G_F\sim 10^{10}G_F$
(from much smaller than weak coupling to much larger than weak
coupling), $g_X=10^{-3}\sim 10^{-1}$ (so we treat the gauge boson mass
$M_X$ as a parameter derived from $G_X$ and $g_X$ and within  the
chosen ranges for these parameters it varies between
1~keV$\lesssim M_X\lesssim 120$ GeV). As for the
sterile mass parameters we allow
$\mst^2=0.01~{\rm eV}^2\sim 10~{\rm eV}^2$ so that the allowed sterile
neutrino mass can be as low as the 95\% CL of the cosmological bounds
on neutrino masses, and can be much larger than the mass suggested by
short baseline anomalies. The sterile neutrino mixing angles 
$\sin^2\theta_{14}$ is varied in a range as
large as $0.003\sim 0.3$ motivated by the latest analysis of the
SBL electron neutrino disappearance data~\cite{Dentler:2018sju}.
In view of the tension between the between SBL $\nu_e$
appearance channel and $\nu_\mu$ results we chose 
$\sin^2\theta_{24}$ to be compatible with the bounds from $\nu_\mu$
disappearance. In order to keep the fit manageable we fix it to be 
30\% of $\sin^2\theta_{14}$. Finally with the existing oscillation
data it is challenging to constrain the mixing between
tau neutrinos and sterile states. So for simplicity we just assume
$\sin\theta_{34}=0$. As we will see later, our results do not depend
on a particular choice of mixing parameters. 

\subsection{BBN}
Let us discuss first  the results obtained from the analysis of BBN
abundances of $^4{\rm He}$ and deuterium.
They have been determined by observations to be~\cite{Izotov:2007ed}
\be
\YP\equiv 4\dfrac{n_{\rm
    He}}{n_b}=0.2465\pm0.0097\,,
\ee
and~\cite{Cooke:2013cba}
\be
\YDP\equiv 10^5\dfrac{n_{\rm D}}{n_{\rm H}}=2.53\pm0.04\,,
\ee
where $n_b$, $n_{\rm He}$, $n_{\rm D}$, $n_{\rm H}$ are the number density
of baryons and helium, deuterium and hydrogen nuclei respectively. To
allow for  a faster confrontation of the model predictions
with BBN data, we use the predicted helium and deuterium abundances
given by Taylor expansions obtained with the PArthENoPE
code~\cite{Pisanti:2007hk}, i.e.
\begin{eqnarray}
  \YP&=&0.2311+0.9502\wb-11.27\wb^2\nonumber\\
  &+&\dNeff'(0.01356+0.008581\wb-0.1810\wb^2)\nonumber\\
  &+&\dNeff^{\prime
    2}(-0.0009795-0.001370\wb+0.01746\wb^2)\,,\\
  \YDP&=&18.754-1534.4\wb+48656\wb^2-552670\wb^3\nonumber\\
  &+&\dNeff'(2.4914-208.11\wb+6760.9\wb^2-78007\wb^3)\nonumber\\
  &+&\dNeff^{\prime
  2}(0.012907-1.3653\wb+37.388\wb^2-267.78\wb^3)\,,
\end{eqnarray}
where $\wb$ is the energy density of baryons today defined as $\wb
\equiv \Omega_bh^2$ and $h\equiv H_0/(100{\rm kms^{-1}Mpc^{-1}})$.
The treatment is similar to that of Planck
Collaboration~\cite{Ade:2015xua}. These theoretical predictions are
subject to errors from neutron lifetime and the interaction rate of
${\rm d(p,} \gamma)^3{\rm He}$ of the order of 
$\sth(\YP)=0.0003$ and $\sth (\YDP) = 0.06$. Since these theoretical errors are
not correlated with the errors from observations, we can add them in
quadrature.

With these data we construct the likelihood
\be
-2\ln{\cal L}(\vec{\omega})=\dfrac{(\YP(\vec{\omega})-\YP^\dat)^2}
{(\sigma^\dat_{\YP})^2+(\sth_{\YP})^2}+\dfrac{(\YDP(\vec{\omega})-\YDP^\dat)^2}
{(\sigma^\dat_{\YDP})^2+(\sth_{\YDP})^2}\,,
\ee
where the relevant model parameters are $\vec{\omega}=(G_X,g_X,\wb,\mst,\ssq)$
and use  MultiNest \cite{Feroz:2007kg,Feroz:2008xx,Feroz:2013hea}
as a Bayesian inference tool.

\begin{figure}[!htb]
\centering
\includegraphics[width=0.75\textwidth]{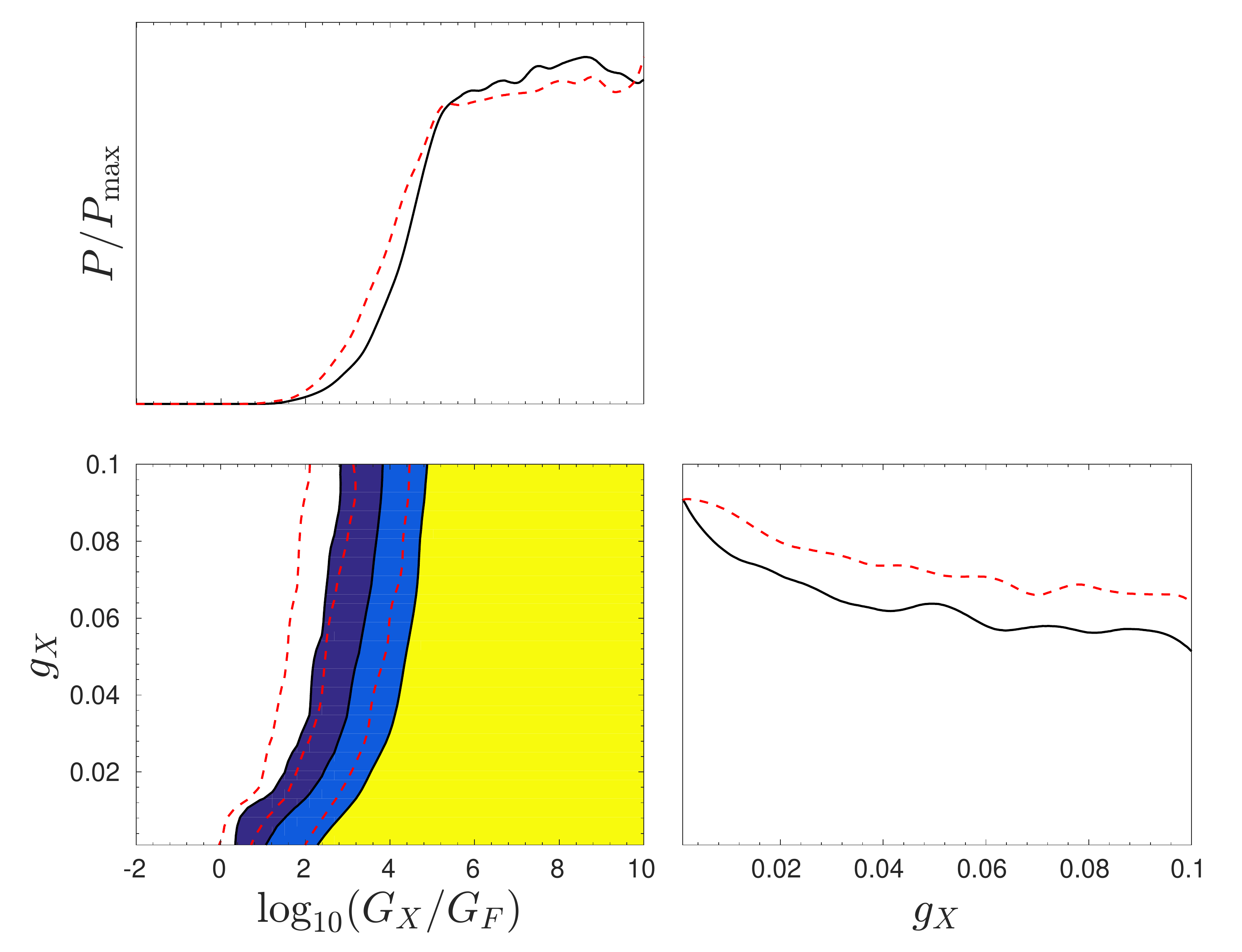}
\caption{Posteriors of $\log_{10}(G_X/G_F)$ and $g_X$ from the fit
  to the BBN abundances of $^4{\rm He}$ and deuterium. The yellow,
  blue and dark blue regions correspond to the $2\sigma$, $3\sigma$
  and $4\sigma$ allowed ranges, respectively.
  {We show for comparison
  the results obtained under the SC assumption of Eq.~\eqref{eq:Neffsol}
  as dashed red lines in the upper left and lower right panels and
void regions in the lower left panel with dashed boundaries representing the corresponding $2\sigma$, $3\sigma$
  and $4\sigma$ allowed ranges, respectively.}}
\label{fig:BBN}
\end{figure}

We assume flat priors on $\log G_X$, $g_X$ and $\wb$ within the range 
$-2\leq \log_{10}(G_X/G_F)\leq 10\,$, $0.001\leq g_X\leq 0.1\,$,
and $0.02153\leq \wb\leq 0.02291\,$,
where the prior on $\wb$ is the 3$\sigma$ range as obtained
from Ref.~\cite{Ade:2015xua}. We also assume gaussian priors
on and $\mst$ and $\ssq$ with $\mst=1.13\pm0.02\eV$ and
$\ssq=0.009\pm0.003$ as motivated by the fit of $\nu_e$ disappearance
data in Ref~\cite{Dentler:2018sju}. {We show for comparison
  the results obtained also under the SC assumption of Eq.~\eqref{eq:Neffsol}
  as dashed lines in the upper left and lower right panels and
void regions in the lower left panel.}
The results are shown in Fig.~\ref{fig:BBN}.

As expected very small $G_X$ is disfavored since it yields $\Neff$
close to 4. Conversely the posterior distribution of $g_X$ is almost flat
meaning this parameter is very mildly constrained by BBN data within
the chosen range. Allowed ranges for the parameters are
{ $g_X\leq 0.065\,(1\sigma)$ and $g_X\leq 0.095\,(2\sigma)${
    ,} and
$\lgGX\geq 6.1 \,(1\sigma)$ and $\lgGX\geq 4.0\, (2\sigma)$.
As seen in the figure if one had obtained $\Neff$ under the assumption of
entropy conservation after decoupling,
the lower bounds  on $\lgGX$ would have been
somewhat {{weaker}} while the allowed range of $g_X$ would be almost the same}.

\subsection{CMB and BAO}
Most sensitivity to the sterile self-interacting model can be
derived from the analysis of the CMB and BAO data.
For concretness we include in the analysis the CMB spectrum from Planck
2015~\cite{Aghanim:2015xee} high multipole
temperature correlation data as well as the low multipole polarization
data (denoted as ``Planck TT+lowP'' in Planck's publications, and we use
``TT'' here for short). We also include 
the measurements of the scale of the baryon
acoustic oscillation (BAO) peaks at different red shifts as measured
in the 6dF Galaxy Survey~\cite{Beutler:2011hx}, the SDSS DR7 main Galaxy
samples~\cite{Ross:2014qpa}, the CMASS~\cite{Anderson:2013zyy} and
LOWZ~\cite{Anderson:2012sa} samples from SDSS DR11 results of BOSS
experiment.

To do the analysis  we interface the output of the QKEs with the
modified CLASS code for the neutrino Boltzmann equations and use
Markov Chain Monte Carlo (MCMC) code Monte Python~\cite{Audren:2012wb}
for parameter inference.  In total we have 10 parameters in this study.
The six cosmological parameters are the same as the base $\lcdm$ model
in Ref.~\cite{Ade:2015xua}, i.e. the baryon
energy density $\wb\equiv \Omega_{\rm b}h^2$, the cold dark matter density
$\wc\equiv\Omega_{\rm cdm}h^2$, the size of sound horizon at
recombination $100\theta_{\rm MC}$, the optical depth to reionisation
$\tau_{\rm reio}$ and the amplitude and tilt of the initial power
spectrum $\ln(10^{10}A_s)$ and $n_s$ respectively.
All six cosmological parameters have
flat priors without upper nor lower limits except $\tre\geq
0.04$.
Besides the six cosmological parameters, we have the four model 
parameters to describe the new scenario: $\lgGX$, $g_X$, $\mst$
and $\ssq$.
Their prior ranges can be found in Table~\ref{tab:priorrange}. We also fix the sum of the neutrino masses of active states to be 0.06~eV.
\begin{table}[h]
\centering \setlength{\tabcolsep}{0.15em} 
           {\renewcommand{\arraystretch}{1.3}
\begin{tabular}{c|c|c|c|c}
Prior&$\lgGX$&$g_X$&$\mst({\rm eV})$ &$\ssq$\\ 
\hline
Broad&$[-2,\,10]$&$[0.001,\,0.065]$&$[0.1,\,3]$&$[0.003\,,0.3]$\\
Narrow&$[-2,\,10]$&$[0.001,\,0.065]$&$1.13\pm 0.02$&$0.009\pm0.003$
\end{tabular}}
\caption{Prior ranges for ``broad prior'' and ``narrow prior''.
All the parameters are flat except $\mst$ and $\ssq$ for narrow
prior which are instead gaussian with the center and width listed above.}
\label{tab:priorrange}
\end{table}

As seen in the table we impose two different priors on $\mst$ and
$\ssq$. One is flat in the range $0.1~{\rm eV}\leq \mst \leq 3~{\rm
  eV}$ and $0.003\leq\ssq\leq 0.3$, which we denote as ``broad
prior''. It was chosen with the aim at studying the information on these
parameters which can be derived from cosmology in the presence of
self-interacting scenario .  The other one is gaussian with
$\mst=1.13\pm0.02~{\rm eV}$ and $\ssq=0.009\pm0.003$ and is the same
as the one we have adopted for the BBN analysis and it aims at
targeting specifically the SBL anomaly. We denote it as ``narrow
prior''. Technically, the narrow prior is imposed by adding a gaussian
likelihood to the data likelihood.

The prior of the parameter $g_X$ is flat in the limited range
$[0.001,\,0.065]$ as motivated by the $1\sigma$ range of BBN
constraint discussed above. We notice that $g_X$ only enters
in the CMB and BAO observables indirectly via its effect on 
the modified energy density and pressure of the
neutrinos at the starting of the background evolution in CLASS.
We have verified that extending its range would not add more freedom to
the analysis while it affects numerical convergence.

\begin{figure}[!htb]
\centering
\includegraphics[width=0.6\textwidth]{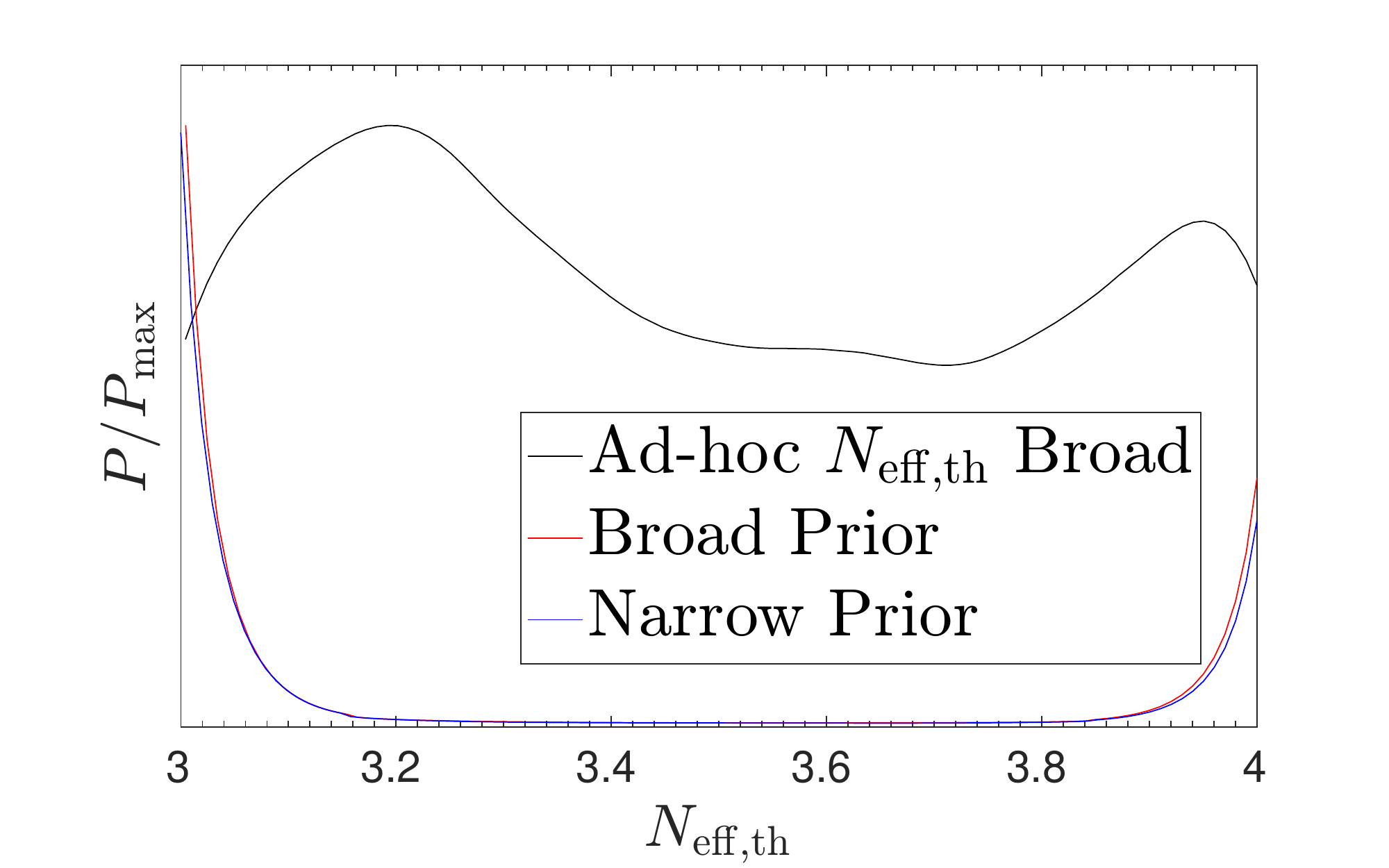}
\caption{Resulting priors on $\Neffth$ for broad prior (red line) and
  narrow prior (blue line). For comparison we also and add-hoc prior
  in which we cut the parameter space of $M_X$ to force a nearly
  flat prior on $\Neffth$. See Appendix \ref{sec:prior} for  details.}
\label{fig:Neffprior}
\end{figure}

As stressed above, in the analysis of CMB and BAO data
$\Neff$ is never used as an input. Still, for a given value of
model parameters we can obtain the value of $\Neff$ at the relevant
temperatures. In this way, given the priors for the four model parameters
we can infer the corresponding prior for $\Neff$
This  is shown  in Fig.~\ref{fig:Neffprior} where we plot
the derived priors of $\Neffth$
corresponding to the two model priors for the relevant parameters.
As $\Neffth$ mainly depends on the effective coupling $G_X$, if  $G_X\gg G_F$,
$\Neffth\simeq 3$; on the other hand, if $G_X\ll G_F$, $\Neffth\simeq 4$.
Therefore the resultant $\Neffth$ of both broad and narrow priors
is peaked at these two extremes and it differs from those extreme values
in a very small range of the input parameter space.
In the figure we also include a modified prior which is nearly flat
in $\Neff$
that we comment upon in the Appendix. {Let us mention 
  that if one uses the assumption of entropy conservation after
  decoupling in the evaluation
of $\Neff$ the prior would be similar in shape but the low peak
would be at $\Neffth=2.7$.}
 
\begin{figure}[!htb]
\centering \includegraphics[width=0.9\textwidth]{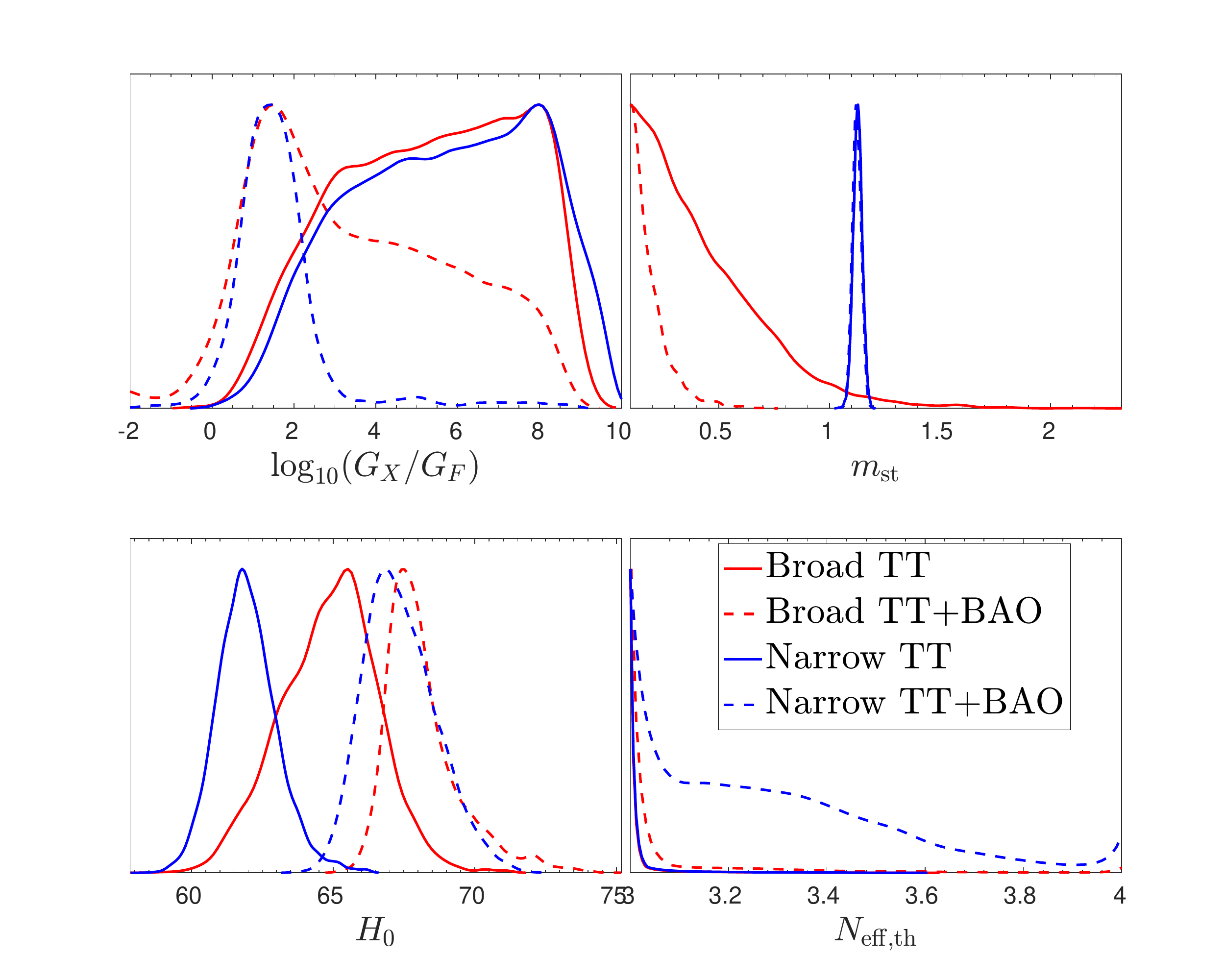}
\caption{Posterior distribution of relevant cosmological and new
  interaction parameters. The solid red, dashed red, solid blue and
  dashed blue lines correspond to {broad} prior with Planck data, {broad}
  prior with Planck+BAO data, narrow prior with Planck data and narrow
  prior with Planck+BAO data, respectively. They are normalized so
  that the maximum probability density is 1. We stress that both $\Neffth$ and
  $H_0$ are derived  parameters.}
\label{fig:post1d}
\end{figure}

We have summarized the results of our analysis in Table~\ref{tab:paramsum} and
Fig.~\ref{fig:post1d} where we give the allowed ranges of the 6+4
parameters and the posterior probability distribution
for the four sterile model parameters respectively.
As expected there is nearly no constraint on
$g_X$ and $\ssq$. The 95\%~CL limit on $\mst$ for broad prior and TT
data is {$\mst<0.95\eV$ . It can be compared with the 95\% limit of
$\mst<0.82\eV$ in terms of the S$\lcdm$ model in
Ref.~\cite{Forastieri:2017oma}. As expected when dropping the assumption of
entropy conservation  after decoupling the limit is
slightly relaxed.} But, still,  even in the presence of self-interactions
with a wide range of couplings a sterile neutrino with mass larger than $1\eV$
is  excluded by more than $2\sigma$. Adding BAO data puts even
tighter constraint on $\mst$ and we find $\mst<0.37\eV$ at 95\%~CL.
{
\begin{table}[h]
\centering \setlength{\tabcolsep}{0.15em} 
           {\renewcommand{\arraystretch}{1.3}
\begin{tabular}{c|cc|cc}
  &\multicolumn{2}{c|}
  {Broad Prior}&\multicolumn{2}{c}{Narrow Prior}\\ \hline
  Parameter&TT&TT+BAO&TT&TT+BAO\\ \hline
  $\wb$&$0.02209^{+0.00022}_{-0.00026}$&$0.02242^{+0.00021}_{-0.00027}$
  &$0.02204\pm 0.00024 $&$0.02282^{+0.00026}_{-0.00030}$\\
  $\wc$&$0.1201\pm0.0022$&$0.1182^{+0.0012}_{-0.0037}$
  &$0.1200^{+0.0022}_{-0.0025}$&$0.1150^{+0.0025}_{-0.0034}$\\
  $100\theta_{\rm MC}$ & $1.04187\pm0.00047$&$1.04200^{+0.00060}_{-0.00043}$
  &$1.04189^{+0.00047}_{-0.00056}$&$1.04220^{+0.00058}_{-0.00055}$\\
  $\As$&$3.093^{+0.033}_{-0.041}$&$3.116\pm 0.041$
  &$3.095\pm 0.037$&$3.170^{+0.041}_{-0.046}$\\
  $n_s$&$0.9592\pm 0.0084$&$0.941^{+0.005}_{-0.010}$&
  $0.9499\pm 0.0078$&$0.9822^{+0.0073}_{-0.012}$\\
  $\tre$&$0.080^{+0.017}_{-0.020}$&$0.093\pm0.020$
  &$0.081\pm 0.019$&$0.123\pm0.020$\\
  \hline
    $\lgGX$&$[1.46,8.90]$&$[-0.03,8.30]$&$[1.74,9.45]$&$[-0.55,5.32]$\\
  $g_X$&-&-&-& $<0.041$\\
  $\mst/{\rm eV}$&$<0.95$&$<0.37$&$1.12\pm0.02$&$1.12\pm0.02$\\
  $\ssq$&-&-&$0.009\pm0.003$&$0.009\pm0.003$\\\hline\hline
  $H_0$&$64.9^{+1.8}_{-1.6}$&$68.16^{+0.58}_{-1.5}$
  &$61.97^{+0.90}_{-1.1}$&$67.4^{+1.0}_{-1.5}$ \\
  $\Neff$ & $<3.21$ & - & $<3.23$ & - \\
\end{tabular}
}
\caption{Allowed ranges for the model parameters for different
  priors and data sets. Cosmological parameters are shown in mean$\pm
  1\sigma$ and the parameters for the new interaction are shown in
  95\%~CL except for $\mst$ and $\ssq$ in narrow priors.
  In entries marked as $-$ the full prior range is allowed.
  We also show
  the corresponding derived ranges for $H_0$ in unit of
  ${\rm km}{\rm s}^{-1}{\rm Mpc}^{-1}$.}
\label{tab:paramsum}
\end{table}
}

We also notice that the interacting $\nu_s$ scenario with large $G_X$
($G_X>10^4G_F$) is preferred over the non-interacting scenario when
considering TT data only. This is expected since small $G_X$ tends to
produce $\Neff\simeq 4$ which is too far away from the favoured value
of 3 to be reconciled with the shift of other cosmological
parameters. What is more surprising is that by adding BAO data, the
favouring of large $G_X$ drops significantly while that of small $G_X$
is lifted.

\begin{figure}[!htb]
\centering \includegraphics[width=0.6\textwidth]{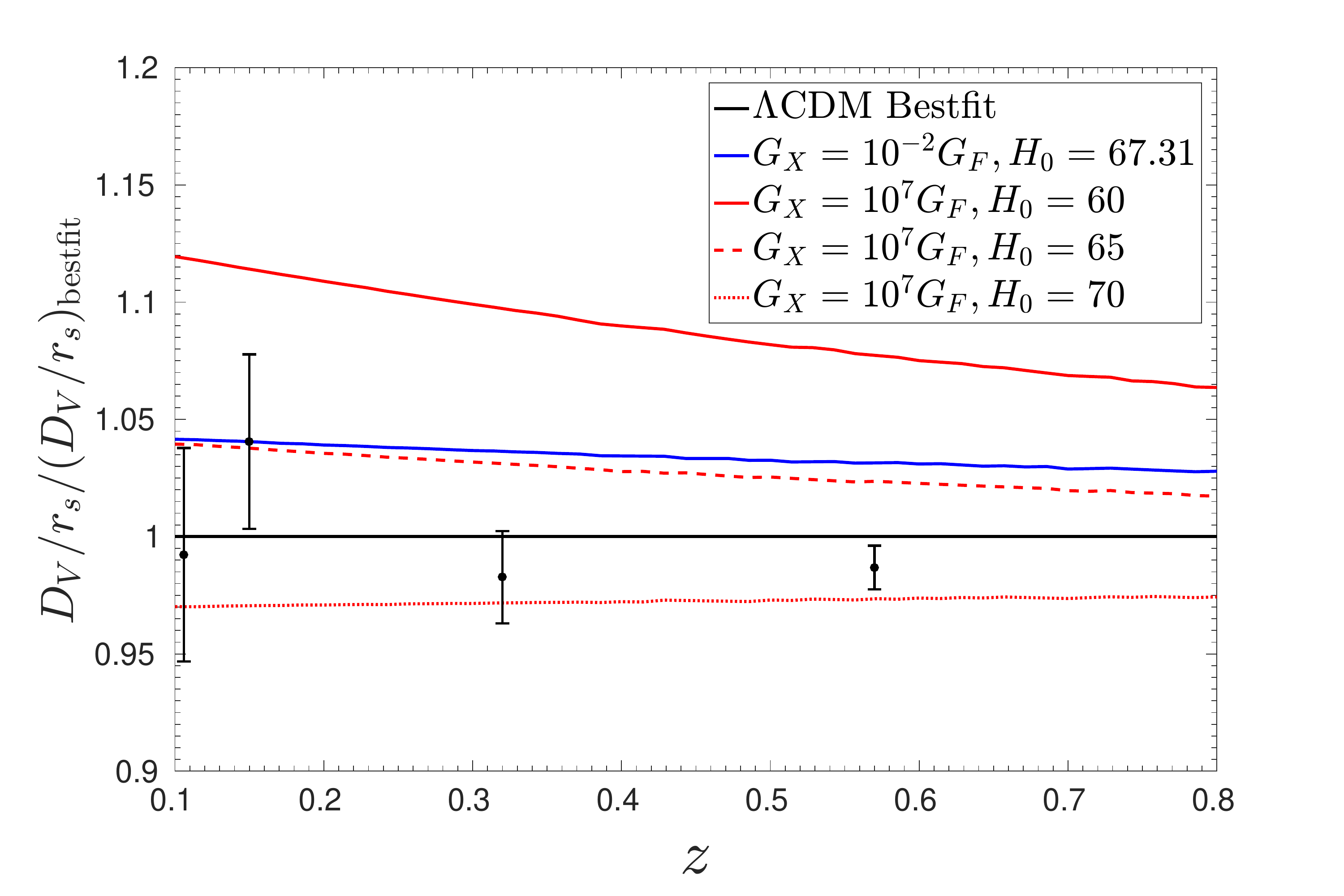}
\caption{$D_V/r_s$ as a function of redshift $z$ normalized to the 
  $\lcdm$ prediction (for the same parameters as  Fig.~\ref{fig:Cl}
  for which $H_0=67.31~{\rm km}{\rm s}^{-1}{\rm Mpc}^{-1}$).
  BAO data and error  bars are shown in black.  
 The colored lines correspond to the interacting $\nu_s$ scenario with
 $g_X=0.1$, $\mst=1~{\rm eV}$, and $\sin\theta_{14}=0.1$ and different values of
 $G_X$ as labeled in the figure. For these curves the other cosmological
 parameters have been fixed to the
 $\lcdm$ best fit except the size of the sound horizon at recombination
 $100\theta_{\rm MC}$ which has
 been adjusted to produce the corresponding values of $H_0$ given in the
 label (in units of ${\rm km}{\rm s}^{-1}{\rm Mpc}^{-1}$).}
\label{fig:BAO}
\end{figure}

To better understand this behaviour we plot in Fig.~\ref{fig:BAO}
the predicted values of the BAO observable 
\be
D_V(z)=\left[(1+z)^2D_A^2(z)\dfrac{cz}{H(z)}\right]^{1/3}\,,
\ee
as a function of redshift.
$D_A(z)$ is the angular diameter distance, $H(z)$Hubble parameter 
and $r_s$ is the comoving sound horizon at the end of the baryon
drag epoch.

Let us stress that in our numerical analysis so far we have used as input
parameters the ten parameters described above, so given a set of
values for the ten parameters, $H_0$ is a derived quantity. For
example for the parameters at the best fit of $\lcdm$ the Hubble
constant comes out to be $H_0=67.31\, {\rm km}{\rm s}^{-1}{\rm
  Mpc}^{-1}$. But when showing the predictions in Fig.~\ref{fig:BAO}
to better control the dependence on $H_0$ we have traded one of the input
parameters,  $100\theta_{\rm MC}$ by $H_0$.
By fixing $\wb$ and $\wc$, $r_s$ is more or less
fixed. But a change in $H_0$ modifies the fraction of dark energy, which changes
the comoving distance back to the baryon drag epoch. Because of this
$D_V$ is also modified.  And as we can see from Fig.~\ref{fig:BAO} for
the strong interacting scenario BAO data favours $H_0$ in between
65~${\rm km}s^{-1}{\rm Mpc}^{-1}$ and 70~${\rm km}s^{-1}{\rm
  Mpc}^{-1}$. Indeed, we see that the prediction for a strongly
interacting $\nu_s$ but with $H_0$ as low as 60~${\rm km}{\rm s}^{-1}{\rm
  Mpc}^{-1}$ is even worse than the case of non-interacting sterile
neutrinos with $H_0=67.31~{\rm km}{\rm s}^{-1}{\rm Mpc}^{-1}$. The
tension cannot be accommodated by varying other cosmological
parameters in the range allowed by CMB data.

\begin{figure}[!htb]
\centering \includegraphics[width=0.6\textwidth]{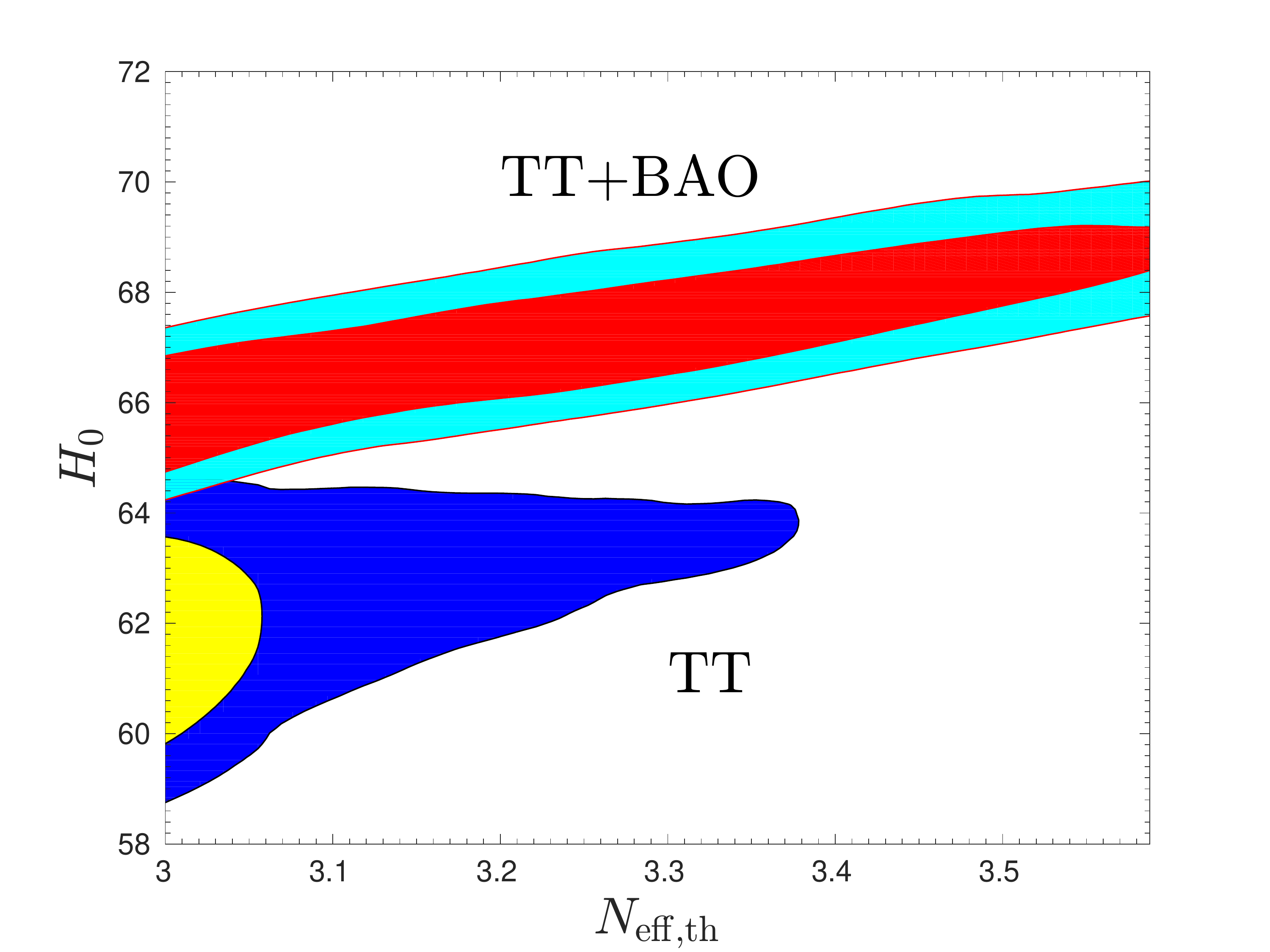}
\caption{2-dimensional posterior correlation of $H_0$ and $\Neffth$ for
  the narrow prior case. The yellow and blue regions show the 68\% and
  95\%~credible intervals using Planck data only. The red and cyan
  regions are the corresponding ones using Planck+BAO data.}
\label{fig:H0Neff}
\end{figure}

However, a small $H_0$ is preferred by TT data, as we can see from
Table~\ref{tab:paramsum} and Fig.~\ref{fig:post1d}. For example when
using the broad (narrow) prior we find that TT data prefers {
$H_0=64.9^{+1.8}_{-1.6}~\uH$ ($H_0=61.97^{+0.90}_{-1.1}~\uH$)}
It is known that there exists a strong correlation between $H_0$ and $\Neff$
which we have shown in Fig.~\ref{fig:H0Neff} by plotting the two-dimensional
posterior allowed regions for those parameters from the analysis with
narrow prior (the corresponding ones for the broad prior are not very
different).  Thus the change in $\Neff$ has to be compensated by the
shift in $H_0$ which explains the small $H_0$ favored by Planck data.
However this small $H_0$ raises a tension with the BAO data. So when
adding BAO to the analysis, a small $G_X$ which predicts large $\Neff$
and large $H_0$ may give a better (if still bad) overall description.
{Let us comment in passing that this tension would have been even worse
  { if }
  one have used the assumption of entropy conservation after neutrino
decoupling to the point that the non-interacting scenario with one
fully thermalized sterile neutrino would give a better fit to TT+BAO
than the strongly interacting case.}

The previous discussion leads us to the question of whether 
the addition of sterile neutrino self-interactions does indeed
help to reconcile eV sterile neutrinos with cosmological observations.
To address this question we have summarized in Table~\ref{tab:minichi2}
the minimum $\chi^2$ obtained in the full parameter space explored
for the different analysis.
{
\begin{table}[h]
\centering \setlength{\tabcolsep}{0.15em} 
{\renewcommand{\arraystretch}{1.3}
\begin{tabular}{c|c|c|c|c|c}
Data&$\lcdm$& Free-$\nu_s$ BP& Free-$\nu_s$ NP &Int-$\nu_s$ BP & Int-$\nu_s$
NP\\ \hline
TT &11261.9 & 9.0 & 18.5 & 1.7 & 6.4 \\
TT+BAO&11266.4 & 7.3 & 32.0 & 1.1 & 22.1
\end{tabular}}
\caption{$\chi^2_{\rm min}$ for various models and data
combinations. The first column show the minimum of
$\lcdm$ model (this $\chi^2_{\rm min}$ is obtained from
the chains of Planck Collaboration available at
\url{http://irsa.ipac.caltech.edu/data/Planck/release_2/ancillary-data/}.),
and others display the shift $\chi^2_{\rm min}$  with respect
to this $\lcdm$  value. ``Free-$\nu_s$ BP'' and
``Free-$\nu_s$ NP'' are models of $\lcdm$ with one
non-interacting sterile neutrino species with $\mst$ and
$\theta_{14}$ priors as in Table~\ref{tab:priorrange} but
with a fixed and very weak interaction ($G_X=10^{-2}G_F$
so effectively for this model $\Neff=4$).}
\label{tab:minichi2}
\end{table}
}
From the table we read that the best fit for the interacting $\nu_s$ model 
with the broad priors is always comparable to the $\lcdm$ best fit, and much
better than the case without new interactions. This is because within the
broad prior there always exists an interaction strength which predicts
$\Neff\simeq 3$ and a reasonable fit can be obtained at the lower limit
of the allowed range of $\mst$.

For the interacting $\nu_s$ model with narrow priors we find that, as
for cosmological data respects, $\lcdm$ is a better fit by { $\Delta
\chi^2_{\rm min}=-6.4$} when using TT data, but still, it provides a
much better description than the non-interacting case.  However,
as read from the table, including BAO data (which are 4 data points)
increases the $\chi^2$ of both interacting and non-interacting
scenarios with the narrow prior by {$\sim$ 15 units} as a consequence of
the tension between the $H_0$ values favoured by CMB and BAO in this
scenario.  Therefore we conclude that self-interactions of
$\nu_s$ have limited power to reconcile the
sterile neutrinos required by the short baseline anomalies when the
BAO information is included.

\section{Conclusions}
\label{sec:conclu}
In this work we have revisited the scenario with  self-interaction
among light sterile neutrinos mediated by a massive gauge boson
proposed to  alleviate the tension between ${\cal O}$(eV) sterile
neutrinos -- motivated by the SBL anomalies--, and the cosmological
bounds on the presence of extra radiation and neutrino masses.  We
have explored a wide range of the model parameters with the goal of
determining if such secret interaction can (or cannot) improve the
description of cosmological data.  

For each point in the model parameter space we  have obtained the effective
number of neutrino species at the time of BBN by solving the QKEs
quantifying the neutrino flavour evolution.
Subsequently we have consistently introduced these results in
the evolution of the density perturbations relevant for predicting
the CMB and  BAO data. In order to do so we have solved the modified
collisional Boltzmann equations for the perturbations accounting also
for the effects of the modified energy density and pressure of the
neutrinos in the background evolution.

With these predictions we have first performed  an analysis of the
BBN data in terms of the primordial abundances of $^4$He and deuterium
which mostly yelds information on the the effective interaction strength
which we find to be bounded  to { $\lgGX\geq 4.0$}  at 95\% CL. 
We have then performed a Bayesian analysis
of CMB and BAO measurements for two different priors on the sterile
neutrino mass and mixings.
By allowing a wide prior for the sterile neutrino mass, we find {$\mst\leq
0.95\eV$ (95\%~CL)} considering Planck data only and $\mst\leq 0.37\eV$
(95\%~CL) using a combination of Planck and BAO data. So the mass
bounds are slightly relaxed compared with that of a non-interacting sterile
neutrino model but a sterile neutrino mass of 1\,eV is still excluded
by more than $2\sigma$ CL. We also performed an analysis by fixing the
sterile neutrino mass and mixing to the values preferred by short
baseline data. Both analysis show that Planck data alone favors
relatively large $G_X$, i.e. the new interaction scenario, while
including BAO information significantly increases the probability of
models with small $G_X$, i.e the non-interacting scenario.  We have
shown how this can be explained by the known degeneracy between $H_0$
and $\Neff$--- the small $\Neff$  leads to small
$H_0$, which is in contradiction with the BAO data. As a consequence
the overall quality of the fit is severely degraded. Furthermore this
is also at odds with the lower bound on the interaction strength
implied by BBN. 

We conclude then that adding the new interaction can alleviate the
tension between eV sterile neutrinos and Planck data, but when
including also the BAO results, the self-interacting sterile neutrino
model cannot provide a consistent global description of the cosmological
observations.

\acknowledgments
We are grateful to Yanliang Shi, Chi-Ting Chiang and Marilena Loverde for very useful discussions. We are also indebted to Stefano Gariazzo for pointing out a typo in the sterile mass prior used in an earlier version of this work. This work is supported by USA-NSF grant PHY-1620628, by EU Networks
FP10 ITN ELUSIVES (H2020-MSCA-ITN-2015-674896) and INVISIBLES-PLUS
(H2020-MSCA-RISE-2015-690575), by MINECO grant FPA2016-76005-C2-1-P
and by Maria de Maetzu program grant MDM-2014-0367 of ICCUB.

\appendix
\section{Appendix}

\subsection{Dependence on the derived prior of $\Neff$}
\label{sec:prior}

The priors for the model parameters used in our analysis
lead to a derived prior for $\Neff$ which is highly peak 
at either 3 or 4 as shown in Fig.~\ref{fig:Neffprior}
and while TT favours models  close to the 3 peak, {BAO disfavours
them especially in the narrow prior case.}
One may wonder then if for models leading to $\Neff$ in the intermediate
region one could find some compromise. This possibility however is not
observed in the posterior distribution of those analysis as a result of
the ``volume effect'' of the priors.

It is important to stress however that the conclusion about the
bad quality of the overall description of the TT+BAO data when
using the narrow prior holds independently of this prior bias because
it is based on the value of the minimum $\chi^2$ of the analysis which is
independent of the shape of prior probability distributions.

Still, to quantify the effect of the bias induced by the model priors in the
derived posterior for the broad prior analysis, we have searched for an
add-hoc prior for the four model parameters which
resulted into a derived prior for $\Neff$ which was as flat as possible.
We found that for this it was best to use $M_X$ and $g_X$ as base
parameters instead of $G_X$ and $g_X$. With a flat prior on
$\log_{10}M_X$ for $M_X$ between {40~MeV} and 1200~MeV and $g_X$  still flat
between 0.001 and 0.065  (so the derived range of
{ $-2\leq \lgGX\leq 4.6$}) and with broad or narrow prior ranges
for $\mst$ and $\theta_{14}$, the derived $\Neff$ prior obtained is that
shown in the corresponding curve in Fig.~\ref{fig:Neffprior}, which,
as seen in the figure, is relatively flat between 3 and 4.  

\begin{figure}[!htb]
\centering \includegraphics[width=0.8\textwidth]{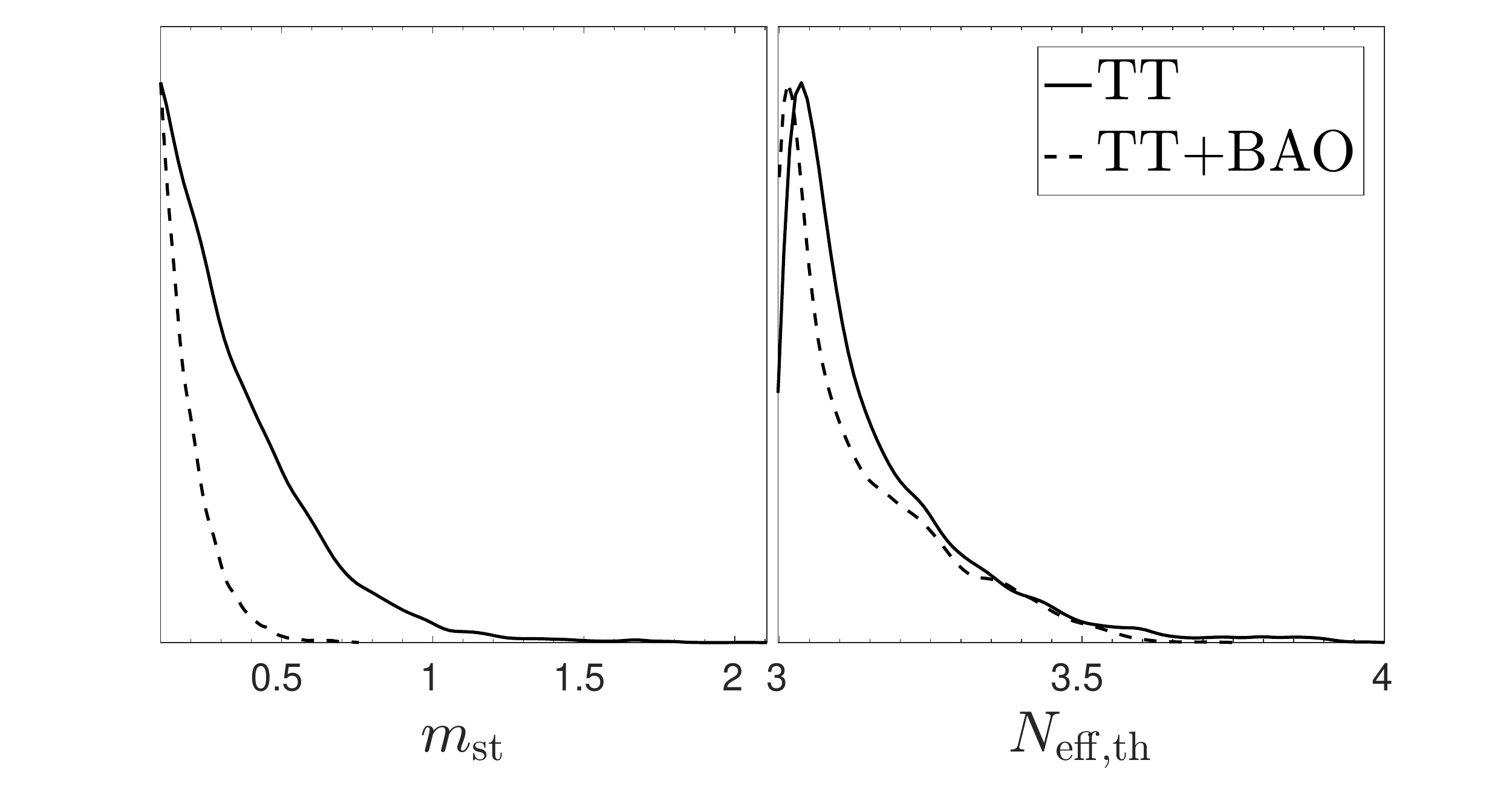}
\caption{Posterior distribution of $\mst$ and $\Neff$ assuming the add-hoc
  $\Neff$ prior and for the data combinations as labeled in the figure.}
\label{fig:MXflat}
\end{figure}

The resulting posterior for $\mst$ and $\Neff$ for the analysis with
this add-hoc broad prior are shown in Fig.~\ref{fig:MXflat}.
{The first thing we notice is that with the ad-hoc prior we do not 
observe the long tale in the $\Neffth$ probability
distribution for the narrow prior observed in Fig.~\ref{fig:post1d}
(and with a  non-negleable probability for the non-interacting case $\Neff=4$)
when including BAO data. We find instead that even when
considering TT+BAO data $\Neff<3.41$ and $\mst<0.37\eV$ at 95\%~CL.
In what respects to the interaction parameters for the add-hoc  prior
we find that the full range of $g_X$ is allowed at this CL while
$40\leq M_X\leq 840$~MeV when using TT data only
(which implies $0.57\leq\lgGX \leq 4.4$) while the  range of
$70\leq M_X\leq 1000$~MeV and $0.47\leq\lgGX\leq 4.1$ is allowed in the
TT+BAO analysis.}

We finish by commenting that when using this ad-hoc quasi-flat $\Neff$
prior we can cross-check the results of our analysis with the
corresponding analysis performed by the Planck collaboration in terms
of an ``effective'' sterile neutrino mass with free $\Neff$
~\cite{Ade:2015xua}.  We find that our results are consistent with
those obtained by Planck collaboration in their analysis (which
includes TT+lensing+BAO data) $\Neff<3.7$ and $m^{\rm eff}_{\nu,{\rm
    sterile}}<0.38\eV$.

\subsection{Validity of $T_\nu<M_X$ approximation}

We have performed our analysis in a relatively broad range of $G_X$ and $g_X$.
They correspond to the gauge boson mass in a range as large as
120~GeV and as small as 1~keV. However when solving the QKEs we always
assume $T_\nu<M_X$ so that the
Lagrangian can be approximated by $4\nu$ effective interaction. One
may question if the approximation is still valid for small
$M_X$.

\begin{figure}[!htb]
\centering \includegraphics[width=0.8\textwidth]{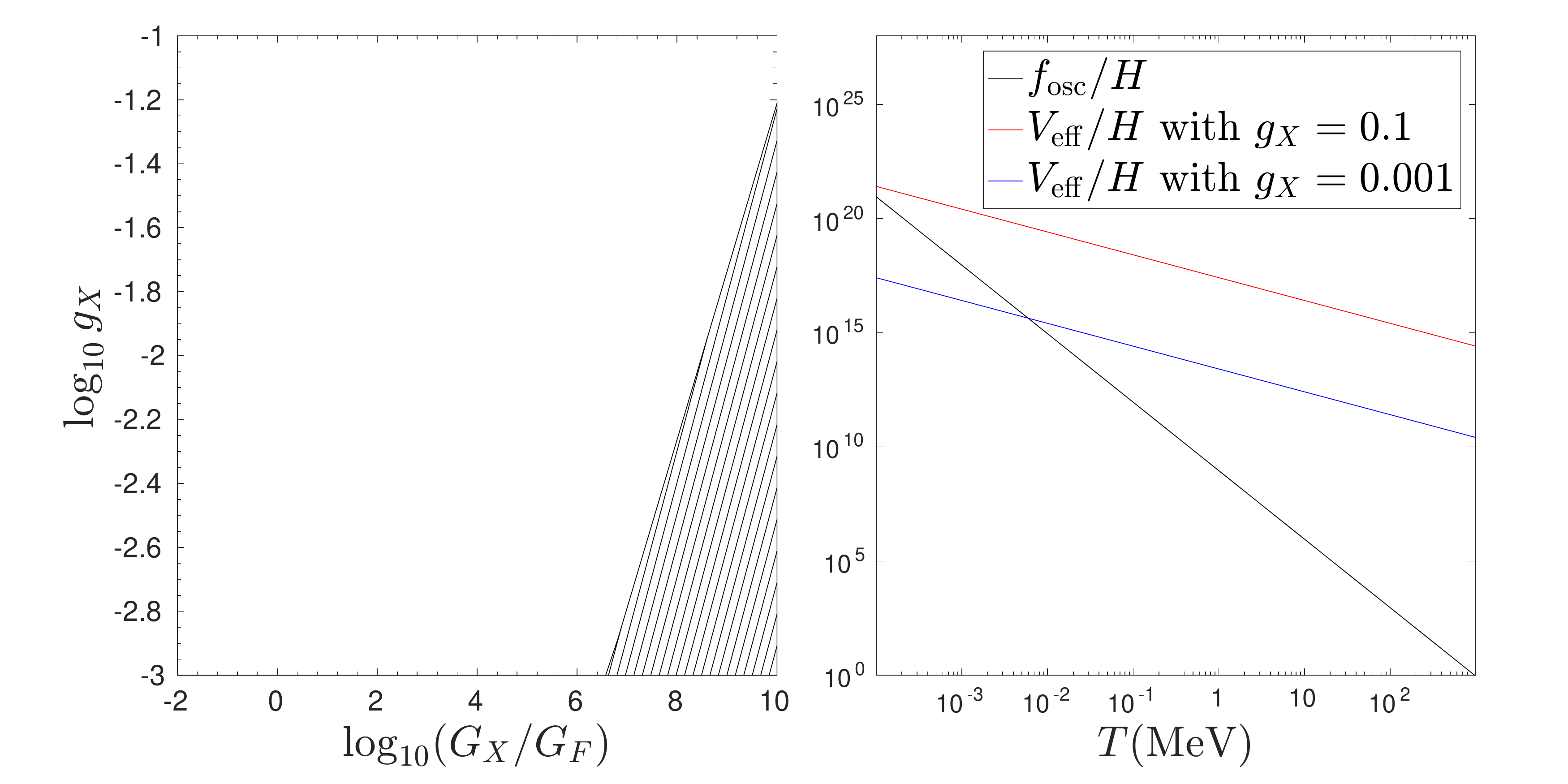}
\caption{Left: Maximum inconsistent region obtained by adding up the
  inconsistent region of all allowed mixing angle and sterile neutrino
  mass. The hatched region may not be well consistent with the
  four-fermion approximation we have adopted in this work. Right: The
  effective potential and oscillation frequency as a function of the
  photon temperature for $g_X=0.1$ (red line) and $g_X=0.001$ (blue line).
  We assume { $T_\nu=(\frac{4}{11})^{1/3}T_\gamma$} and $\mst=1\eV$.}
\label{fig:saferegion}
\end{figure}

To estimate the parameter region in which our approximation is
not valid we look for the parameters for which more than 50\% of the 
sterile neutrinos are produced at $T_\nu>M_X$. 
Technically we require $\rho_{ss}$ not to have reached
0.5 when neutrino temperature drops below $M_X$.
We show in the left panel of Fig.~\ref{fig:saferegion} the
inconsistent region of $G_X$ and $g_X$ while allowing mixing angle and
$\mst$ to vary within the solution limits. In other words, this is the
maximum inconsistent region for $G_X$ and $g_X$.
Recalling the definition of
$G_X$ ($\frac{G_X}{\sqrt{2}}=\frac{g_X^2}{8M_X^2}$), the boundary
line of the region corresponds to a constant $M_X\simeq 64~{\rm keV}$.
Inside the region $M_X$ is smaller than this value. 

From the solutions of QKEs we find that for  all the parameters in the
inconsistent region we always obtain  {$\Neff\simeq 3$}, regardless of the choice
of $\mst$ and $\ssq$. We argue next that despite the approximation does
not hold, the solution obtained is still valid. 

To qualitatively confirm this, we look at what would be the effective
potential in case $T_\nu>M_X$ where $\Veff\simeq \frac{g_X^2T_\nu^2}{8E}$
\cite{Chu:2015ipa}  and compare it with the oscillation
frequency and the Hubble parameter.
This is shown in the right panel of Fig.~\ref{fig:saferegion}
(as a qualitative estimate we average $1/E$ over the phase space distribution
of neutrinos and replace it by $0.456T_\nu$ in the plotted $\Veff$).
As seen in the figure, in the full range of $g_X$ we use,  the
effective potential is always much larger than the oscillation
frequency for temperatures above 1 MeV. This again leads to the suppression of
the in-medium mixing angle in Eq.~\ref{eq:sinthetam} and  the late
production of sterile neutrinos which leads to {$\Neff\simeq 3$}.
Notice that the maximum $M_X$ of 64~keV is much smaller than 1~MeV
so for these curves it is always the case that $T_\nu>M_X$ 
We conclude that our solutions do still hold
even in the region where $T_\nu<M_X$ approximation fails.

One caveat of the above argument is the possible effects associated
with the direct production of the massive gauge boson if light enough.
In the region of parameters we consider this is only expected to
happen well after neutrino decoupling and therefore cannot lead
to an increase of $\Neff$. 
As a matter of fact, a massive gauge boson model in the $T_\nu\gg M_X$
limit is qualitatively equivalent to the massless pseudoscalar model
discussed in Ref.~\cite{Archidiacono:2014nda} where the QKEs describing
two-neutrino oscillation shows $g_X$ has to be smaller than $10^{-5}$
to achieve $\Neff>3$.  This tells us that in our parameter range, 
the condition $T_\nu\gg M_X$ implies that the thermalizaiton 
between active and sterile neutrinos has never happened.
In other words in our scenario $\Neff>3$ can only be realized
when $T_\nu<M_X$ above 1~MeV and the effective coupling $G_X$ is small enough.

\bibliographystyle{JHEP}
\bibliography{references}

\end{document}